\documentstyle[multicol,aps,epsf]{revtex}

\title{Phase diagrams of classical spin fluids: the influence of
       an external magnetic field \\ on the liquid-gas transition}

\author{W. Fenz,$^1$ R. Folk,$^1$ I. M. Mryglod,$^{1,2}$ and
        I. P. Omelyan$^{1,2}$}

\address{$^1$Institute for Theoretical Physics, Linz University,
         A-4040 Linz, Austria}
\address{$^2$Institute for Condensed Matter Physics,
         1 Svientsitskii Street, UA-79011 Lviv, Ukraine}

\date{\today}

\begin{document}

\maketitle

\begin{abstract}

The influence of an external magnetic field on the liquid-gas
phase transition in Ising, XY, and Heisenberg spin fluid models
is studied using a modified mean field theory and Gibbs ensemble
Monte Carlo simulations. It is demonstrated that the theory is
able to reproduce quantitatively all characteristic features of
the field dependence of the critical temperature $T_{\rm c}(H)$
for all the three models. These features include a monotonic
decrease of $T_{\rm c}$ with rising $H$ in the case of the Ising
fluid as well as a more complicated nonmonotonic behavior for
the XY and Heisenberg models. The nonmonotonicity consists in a
decrease of $T_{\rm c}$ with increasing $H$ at weak external
fields, an increase of $T_{\rm c}$ with rising $H$ in the strong
field regime, and the existence of a minimum in $T_{\rm c}(H)$
at intermediate values of $H$. Analytical expressions for $T_{\rm
c}(H)$ in the large field limit are presented as well. The
magnetic para-ferro phase transition is also considered in
simulations and described within the mean field theory.

\vspace{12pt}

\noindent
PACS number(s): 05.70.Fh, 64.60.-i, 64.70.Fx, 75.50.Mm

\end{abstract}

\vspace{15pt}

\begin{multicols}{2}

\section{Introduction}

The investigation of continuum fluid models with coupled translational
and spin degrees of freedom is of current theoretical interest
\cite{NijWe,Mryglod,Fenz}. The importance of such models lies in their
property to display a rich variety of transitions between solid, liquid,
and gas, as well as magnetic ordered and disordered phases, which may
occur in real systems. The liquid-gas and magnetic para-ferro phase
transitions in spin fluids were studied previously by the mean field
theory \cite{Fenz,Frankel,HemImb77,Tavares,Schinagl}, the method of
integral equations \cite{Lom,Sokolovska,Lados,Lado,Sokolovskii,Lomba},
and Monte Carlo (MC) simulation techniques \cite{Mryglod,Lom,Lado,%
Nijmeijer,Weis,Parola,Ferreira,Nielaba}. The theoretical studies dealt
mainly with spatially one-($d=1$) and three-($d=3$) dimensional Ising
($n=1$) as well as three-dimensional Heisenberg ($n=3$) fluids (here
$n$ denotes the spin dimensionality). In computer experiment, the
magnetic transition was investigated for the Heisenberg model
\cite{Mryglod,Nijmeijer} as well as for three- \cite{Parola} and
two-($d=2$) \cite{Ferreira} dimensional Ising ($n=1$) fluids using
canonical MC simulations. The combined canonical and Gibbs ensemble
MC (GEMC) simulations were performed for a Heisenberg system ($d=3$,
$n=3$) to determine both the magnetic and liquid-gas transitions
\cite{Lom,Lado,Weis}.

In most of the previous works, the liquid-gas coexistence was evaluated
in the absence ($H=0$) of an external magnetic field. Only a few papers
\cite{Schinagl,Lado,Sokolovskii} were devoted to theoretical study of
the fluid behavior at $H \ne 0$. It was found that for systems of hard
spheres carrying Ising spins, an external magnetic field decreases the
temperature $T_{\rm c}$ of the gas-liquid critical point \cite{Schinagl}.
On the other hand, the presence of Heisenberg spins can lead to the
inverse effect at strong enough fields \cite{Lado,Sokolovskii}. As a
result, a nonmonotonic behavior of $T_{\rm c}$ may arise in Heisenberg
fluids due to a subtle interplay between the translational and spin
degrees of freedom \cite{Sokolovskii}. To our knowledge, no confirmation
of the nonmonotonicity in $T_{\rm c}(H)$ has been given within computer
experiment for continuum fluid models of spin systems. In the only work
\cite{Lado} done on GEMC simulations at $H \ne 0$ for a Heisenberg fluid,
it has been concluded that the application of an external field increases
the gas-liquid critical temperature $T_{\rm c}$. But this conclusion has
been made on the basis of results corresponding to just one finite
(sufficiently large) value of $H$.

It is worth mentioning also that, as far as we are aware, no theoretical
calculations and computer simulations of $T_{\rm c}(H)$ have been
performed for the planar ($n=2$, $d=3$) XY spin fluid model and no
simulations on liquid-gas coexistence in the presence of an external
field have been reported for the three-dimensional Ising fluid. Note
that we are considering genuine fluid models in which spatial positions
of spins are changed continuously (contrary to simplified so-called
lattice gas schemes \cite{Kawasaki,Sokolovski,Romano} where spins are
allowed to occupy only positions belonging to sites of a chosen lattice).
Moreover, all the works dealt with nonmagnetic repulsion interactions
in the form of the simplest hard-sphere potential exclusively. In
addition, the magnetic interactions were truncated, as a rule, at
some finite interatomic separation, without taking into account long
range corrections. The question of how these restrictions impact the
behavior of $T_{\rm c}(H)$ has not been considered as well.

In this paper we present a comprehensive study of the influence of
an external magnetic field on the liquid-gas coexistence properties
of fluid models with Ising, XY, and Heisenberg spin interactions.
The corresponding phase diagrams are calculated in the whole region
of varying $H$ with the help of the GEMC simulation technique and
a modified version of the mean field theory. As is shown, a good
agreement between the MC data and theoretical results for $T_{\rm
c}$ can be achieved for all the models and for any value $H$ of
the external field, including the limit $H \to \infty$. At $H=0$,
it is demonstrated that the mean field theory predicts a tricritical
point independent on the dimension $n$ of the magnetic order
parameter. The simulation results for the Ising model agree with
this topology, whereas for the XY and Heisenberg fluids a critical
end point, beside a gas-liquid critical point in the ferromagnetic
phase, is indicated.

\vspace{6pt}

\section{Mean field theory}

\subsection{Models}

Let us consider three models of magnetic fluids for spatial dimension
$d=3$ with spin interactions of Ising ($n=1$), planar XY ($n=2$), and
Heisenberg ($n=3$) types. Within all these models, the total potential
energy of the $N$-particle system can be cast in the form
\begin{equation}
U=\frac{1}{2} \sum_{i \ne j}^{N} \Big[ \varphi(r_{ij}) -
J(r_{ij}) \, {\bf s}_i \! \cdot {\bf s}_j \Big] - {\bf H} \cdot
\sum_{i=1}^{N} {\bf s}_i \, ,
\end{equation}
where ${\bf r}_i = ({r_i}_x,{r_i}_y,{r_i}_z)$ denotes the spatial
coordinate, ${\bf s}_i$ is the $n$-dimensional spin vector [i.e.,
$({s_i}_x,0,0)$, $({s_i}_x,{s_i}_y,0)$, or $({s_i}_x, {s_i}_y,
{s_i}_z$) for $n=1$, 2, or 3, respectively] of unit length
($|{\bf s}_i| = 1$), and $r_{ij}=|{\bf r}_i-{\bf r}_j|$. For
convenience, the homogeneous external magnetic field ${\bf H}=
(H,0,0)$ is directed along axis $X$ of the laboratory system of
coordinates. The Yukawa
function
\begin{equation}
J(r) = \frac{\epsilon \sigma}{r} \exp \bigg[
\frac{\sigma-r} {\sigma} \bigg]
\end{equation}
is used to describe the internal magnetic interactions, where $\sigma$
and $\epsilon$ relate to the size of particles and coupling constant,
respectively. The nonmagnetic interaction $\varphi$ between particles
can be modeled by the hard-sphere (HS)
\begin{equation}
\varphi_{\rm HS}(r) = \left\{
\begin{array}{rl}
\displaystyle
\infty \, , & \ \ \ r < \sigma \, , \\ [12pt]
0 \, \, , & \ \ \ r \ge \sigma
\end{array}
\right.
\end{equation}
or soft-core (SC)
\begin{equation}
\varphi_{\rm SC}(r) = \left\{
\begin{array}{cc}
\displaystyle
4 \epsilon \bigg[ \Big( \frac{\sigma}{r}
\Big)^{12} - \Big( \frac{\sigma}{r} \Big)^6 \bigg] +
\epsilon \, , & \ \ \ r < \sqrt[6]{2} \sigma \, , \\ [12pt]
0 \, , & \ \ r \ge \sqrt[6]{2} \sigma
\end{array}
\right.
\end{equation}
repulsion potentials. Note that the attraction between particles is
formed exclusively due to ferromagnetic interactions. This corresponds
to a so-called ``ideal'' class of spin fluids, where the attractive
part of nonmagnetic interactions is absent (see comments at the end
of the paper).

\subsection{Equations of state}

Following the spirit of works \cite{Tavares,Evans}, the Gibbs free energy
per particle corresponding to Hamiltonian (1) can be presented as
\begin{equation}
{\cal F} = {\cal F}_\varphi + \frac{\langle \ln f({\bf s}) \rangle}
{\beta} + \int_0^1 {\rm d} \alpha {\cal F}_{J_\alpha } - {\bf H} \cdot
{\bf m} \, .
\end{equation}
Here ${\cal F}_\varphi$ is the free energy of the reference system
(in the absence of magnetic interactions and external fields), $f({\bf
s})$ relates to the normalized single-particle function describing the
distribution of spins in orientational space, $\beta^{-1}=k_{\rm B} T$
is the temperature with $k_{\rm B}$ being the Boltzmann's constant,
$\langle \ \rangle$ denotes the statistical averaging, and
\begin{equation}
{\bf m} = \langle {\bf s} \rangle =
\frac{1}{N} \bigg \langle \sum_{i=1}^N {\bf s}_i \bigg
\rangle = \int {\bf s} f({\bf s}) {\rm d} {\bf s}
\end{equation}
defines the magnetization of the system. The contribution to the free
energy caused by spin interactions with the parameterized Yukawa
function $J_\alpha(r)$ can be written in the form
\begin{equation}
{\cal F}_{J_\alpha}\!\!=\!
- \frac{\rho}{2} \int \! \! g_\alpha(r,{\bf s}_1,{\bf s}_2)
\frac{{\rm d} J_\alpha(r)}{{\rm d \alpha}} \, {\bf s}_1 \! \cdot
{\bf s}_2 f({\bf s}_1) f({\bf s}_2) {\rm d}
{\bf r} {\rm d} {\bf s}_1 {\rm d} {\bf s}_2 ,
\nonumber
\end{equation}
where $\rho=N/V$ is the number density with $V$ being the volume, and
$g_\alpha(r,{\bf s}_1,{\bf s}_2)$ introduces the pair distribution
function of a system corresponding to $\alpha$-switched on magnetic
interactions with $J_\alpha=0$ at $\alpha=0$ and $J_\alpha=J(r)$ at
$\alpha=1$.

Equation (5) formally leads to exact results but requires the knowledge
of function $g_\alpha(r,{\bf s}_1,{\bf s}_2)$ for each intermediate
states $0 \le \alpha \le 1$. Since, in general, this function cannot
be determined exactly, some approximations are needed to calculate
${\cal F}$. Within the mean field (MF) approximation it is assumed
that $g_\alpha(r,{\bf s}_1,{\bf s}_2)$ does not depend on $\alpha$
and accepts the form of the pair distribution function $g_\varphi(r)$
of the reference system. The latter function can further be
approximated by its values in the low density regime, $g_\varphi(r)
\approx \exp[-\beta \varphi(r)]$. Then in view of Eqs.~(2) and (6),
the integrations in (5) and (7) can be performed explicitly. This
results in
\begin{equation}
{\cal F} = {\cal F}_\varphi + k_{\rm B} T \int {\rm d} {\bf s} f({\bf
s}) \ln f({\bf s}) - \frac{1}{2} a \rho m^2 - {\bf H} \cdot {\bf m} \, ,
\end{equation}
where
\begin{equation}
a = 4 \pi \int_0^\infty g_\varphi(r) J(r) r^2 {\rm d r} =
8 \gamma(T) \pi \epsilon \sigma^3
\end{equation}
is the magnetic interaction strength, and the multiplier
\begin{equation}
\gamma(T) = \frac{\int_0^\infty \exp[-\beta \varphi(r)] J(r) r^2
{\rm d} r} {\int_\sigma^\infty J(r) r^2 {\rm d} r}
\end{equation}
takes into account the softness of nonmagnetic repulsion potential
$\varphi$.

Considering the free energy (8) as a functional of $f({\bf s})$, it can
be shown that the minimum of ${\cal F}$ is achieved at
\begin{equation}
f({\bf s})=\frac{\exp(\beta {\bf h} \cdot {\bf s})}
{\int \exp(\beta {\bf h} \cdot {\bf s}) \, {\rm d} {\bf s}} \, ,
\end{equation}
where

\vspace{-9pt}

\begin{equation}
{\bf h} = {\bf H} + a \rho {\bf m}
\end{equation}
can be treated as the effective field, consisting of the
external term ${\bf H}$ and averaged internal contribution $a \rho
{\bf m}$. Eq.~(11) defines, therefore, the equilibrium single-particle
distribution function in the MF approximation. Then, taking into account
the fact that the vector ${\bf m}$ is parallel to ${\bf H}$, the
right-hand side of Eq.~(6) can be integrated in quadratures. The
result is
\begin{equation}
m = \left\{
\begin{array}{lc}
\tanh\left(\frac{H + a \rho  m}{k_{\rm B} T}\right)
\, , \ \ \ & n = 1 \, , \nonumber \\ [12pt]
\displaystyle
\frac{I_1\left(\frac{H + a \rho  m}{k_{\rm B} T}\right)}
{I_0\left(\frac{H + a \rho  m}{k_{\rm B} T}\right)}
\, , \ \ \ & n = 2 \, , \\ [23pt]
\coth\left(\frac{H + a \rho  m}{k_{\rm B} T}\right)
- \frac{k_{\rm B} T}{H + a \rho  m}
\, , \ \ \ & n = 3 \, , \nonumber
\end{array}
\right.
\end{equation}
where $I_l(x)=\frac{1}{\pi} \int_0^\pi {\rm e}^{x \cos{\psi}} \cos(l
\psi) {\rm d} \psi$ denotes the modified Bessel function of the first
kind and order $l$. Relation (13) represents the magnetic equation of
state (MES) of the system. Note that the form of this equation depends
on the number $n$ of components of the magnetic order parameter ${\bf
m}$.

The pressure equation of state (PES) can readily be obtained by partially
differentiating Eq.~(8) with respect to $\rho$, using the thermodynamic
relation $P=\rho^2 (\partial {\cal F}/\partial \rho)_{T,H}$ with Eqs.~(11)
and (12). As a consequence, one finds that the total pressure is the sum
of two terms,
\begin{equation}
P=P_\varphi-\frac{1}{2} a \rho^2 m^2 \, ,
\end{equation}
namely, the pressure $P_\varphi=(\rho^2 \partial {\cal F}_\varphi/
\partial \rho)_{T,H}$ corresponding to the reference system and the part
coming from the magnetization. For the HS reference system (3) we use
the quasiexact Carnahan-Starling relation \cite{CarnStar69}
\begin{equation}
P_\varphi(\rho,T)= \rho k_{\rm B} T(1+\eta+\eta^2-\eta^3)(1-\eta)^{-3}
\, ,
\end{equation}
with $\eta=\pi\rho\sigma^3/6=\pi \rho^\ast/6$ being the packing
fraction. In the case of a SC potential (4), the softness of $\varphi$
is taken into account by replacing in Eq.~(15) the HS diameter
$\sigma$ by its SC counterpart $\sigma_\varphi$. The latter quantity
can be determined by requiring the second virial coefficients related
to the HS system with the particle's diameter $\sigma_\varphi$ and
the SC system with the real potential $\varphi$ to be equal between
themselves. This leads to
\begin{equation}
\sigma_\varphi(T) = \xi^{1/3}(T) \, \sigma \, ,
\end{equation}
where
\begin{equation}
\xi(T) = \frac{3}{\sigma^3} \int_0^\infty
\Big(1-\exp[-\beta \varphi(r)]\Big) r^2 dr \, .
\end{equation}
Then the SC pressure can be obtained using Eq.~(15) with $\eta \equiv
\eta_\varphi=\pi \rho \sigma_\varphi^3/6=\xi(T) \pi \rho^*/6$. This is
justified by the fact that the SC potential (4) is close enough to
the HS function (3) ($\varphi_{\rm SC}(r)$ increases rapidly to
infinity with decreasing $r$ in the range $r<\sigma$, whereas it
quickly tends to zero at $r > \sigma$).

The relations (9), (13) and (14) constitute the main results of the
MF theory. In the case of the HS potential (4) (when $g_\varphi(r)=1$
for $r \ge \sigma$ and $g_\varphi (r)=0$ at $r < \sigma$ and, thus,
$\gamma=\xi=1$, see Eqs.~(10) and (17)), they coincide completely
with those obtained earlier \cite{HemImb77,Tavares} (for $n=1$ and
3). Our expressions are more general, since they do not restrict
us to the HS convention only, but also are directly applicable for
more realistic SC magnetic systems (including the case $n=2$).

\subsection{Phase separations}

Analyzing the MES (13) at $H=0$, it can be shown that nontrivial
(nonzero) solutions in the magnetic ordering parameter $m$ take place
for temperatures lower than the Curie temperature $T_\lambda = a
\rho/(n k_{\rm B})$. In the dimensionless representation $\rho^\ast=
\rho \sigma^3$ and $T^\ast=k_{\rm B} T/\epsilon$, the magnetic phase
transition curve reads $T^\ast_\lambda = 8 \pi \rho^\ast \gamma(T^
\ast_\lambda)/n$. Since, in general, the function $\gamma(T)$ may
depend on temperature in a characteristic way, the last equality
represents a nonlinear equation which should be solved with respect
to $T^\ast_\lambda$ at fixed $\rho^\ast$. In the case of SC potential
(4), the computations show that the deviations of $\gamma(T)$ from
unity do not exceed about 3\% in a wide temperature range of $0.3 <
T^\ast <6$. For this reason we can put $\gamma(T)=1$ without loss of
precision (at least in the range mentioned above). Hence, the Curie
temperature is found analytically, $T^\ast_\lambda = 8 \pi \rho^\ast/n$.
It linearly depends on the density and is inversely proportional to the
number of spin components. Near the Curie line at $T \lesssim T_\lambda$,
the MES can be expanded in a series with respect to the deviation
$t=(T_\lambda-T)/T_\lambda$, yielding
\begin{equation}
\lim_{t \to +0} m^2 \frac{T_\lambda}{T_\lambda-T} = c \, ,
\end{equation}
where $c$ is a constant depending on the spin dimensionality $n$ (see
Table~I).

In order to get the liquid-gas critical point $T_{\rm c}$, one has to
look where the inverse compressibility goes to zero. The liquid-gas
phase transition occurs on the Curie line, so that in view of Eqs.~(14),
(15), and (18) one finds at $T=T_\lambda$ that
\begin{equation}
\beta \left(\frac{\partial P}{\partial \rho}\right)_{T,H} =
\frac{\partial}{\partial \eta} \left[ \eta
\frac{1+\eta+\eta^2-\eta^3}{(1-\eta)^3} \right]
- \frac{c n}{2} = 0 \, .
\end{equation}
Thus solving Eq.~(19) with respect to $\eta$ yields a tricritical point
at $\rho^\ast_{\rm t}=6 \eta_{\rm t}/\pi$ and $T^\ast_{\rm t}=8 \pi
\rho^\ast_{\rm t}/n=48 \eta_{\rm t}/n$. Within the HS version of the MF
theory, the solutions $\eta_{\rm t}$ allow to be presented analytically:
$\eta_{\rm t}=1+[\sqrt{q}-(432/\sqrt{q}-72-q)^{1/2}]/6$ with $q=
6(4p)^{1/3}-24-96(2/p)^{1/3}$ and $p=83+3\sqrt{993}$ for $n=1$,
$\eta_{\rm t}=1+\sqrt{q}-(1/\sqrt{q}-1-q)^{1/2}$ with $q=(p^{1/3}-
2-17/p^{1/3})/6$ and $p=82+3\sqrt{1293}$ for $n=2$, and $\eta_{\rm
t}=1+[2\sqrt{q}-(72/\sqrt{q}-24-4q)^{1/2}]/6$ with $q=(p/2)^{1/3}-
26(2/p)^{1/3}-2$ and $p=245+9\sqrt{1609}$ for $n=3$. The values
for $\rho^\ast_{\rm t}$ and $T^\ast_{\rm t}$ with three digits
after the decimal point are collected in Table~I and marked by
the superscript HS. In the case of the SC version, we should
replace (after taking the partial derivative) the packing fraction
$\eta$ entering in Eq.~(19) by its effective value $\eta_\varphi=
\xi(T) \pi \rho^*/6$. It can be shown that for the SC potential
defined by Eq.~(4), the multiplier $\xi^{1/3}(T)$ decreases
monotonically from 1.05 to 0.94 with increasing the temperature
in the interval $T^\ast \in [0.3, 6]$. This behavior has a simple
physical meaning, namely, with increasing $T$ the particle
can approach to one another more closely due to the increase of
their thermal velocities. As a consequence, the effective diameter
$\sigma_\varphi=\xi^{1/3}(T) \sigma$ will decrease. In such a
situation (when $\xi \equiv \xi(T^\ast)$ with $T^\ast=8 \pi
\rho^\ast/n$), Eq.~(19) transforms into a complicated nonlinear
equation in $\rho^\ast$ and must be solved numerically (the
integration (17) has been carried out numerically too). The
results of these computations are shown in Table~I as well
and marked by the superscript SC.

\end{multicols}

\begin{table}[h]
\caption{\label{tab1}{Mean field theory results in $d=3$}}
\begin{center}
\begin{tabular}{cccccccc}
& & & & & \\
Model   & $n$    &  $c$  & $\displaystyle
T^\ast_\lambda$  &
${\rho^\ast_t}^{\rm HS}$ &
${T^\ast_t}^{\rm HS}$ &
${\rho^\ast_t}^{\rm SC}$ &
${T^\ast_t}^{\rm SC}$
\\
 $$      &     &                    &        & & \\
\hline
 $$      &     &                    &        & & \\
  Ising  & 1 &   3  & $8 \pi \rho^\ast$ &
0.098 &
2.462 &
0.104 &
2.621
\\
$$       &     &                     &       & & \\
  XY     & 2 &   2  & $4 \pi \rho^\ast$ &
0.169 &
2.119 &
0.176 &
2.207
\\
 $$      &     &                     &       & & \\
\ Heisenberg \ & 3 & $\displaystyle\frac{5}{3}$ &
$\displaystyle \frac{8 \pi}{3} \rho^\ast$ &
0.224 &
1.876 &
0.229 &
1.920
\\
& & & & & \\
\end{tabular}
\end{center}
\end{table}

\begin{multicols}{2}

In the presence of an external field (i.e., when $H \ne 0$) there is a
liquid-gas phase transition curve ending in a critical point $T_{\rm
c}$. The critical temperature $T_{\rm c}$ and density $\rho_{\rm c}$
can be obtained numerically by solving the following system of two
equations
\begin{equation}
\left(\frac{\partial P}{\partial \rho}\right)_{T,H} = 0 \, , \ \ \ \ \ \ \
\left(\frac{\partial^2 P}{\partial \rho^2}\right)_{T,H} = 0 \, ,
\end{equation}
where, in view of Eqs.~(13) and (14), the pressure $P(\rho,T,m)=P(\rho,T,
m(\rho,T,H)) \equiv P(\rho,T,H)$ should be considered as a function of
$\rho$, $T$, and $H$. Therefore, the solutions $T_{\rm c}(H)$ and
$\rho_{\rm c}(H)$ to system (20) must be found self-consistently with
the solution $m \equiv m(\rho,T,H)$ to nonlinear equation (13). The
latter equation requires to be handled numerically as well. In the
case $n=2$, we have used a representation of the modified Bessel
functions (appearing in Eq.~(13)) in the form of the infinite
series $I_l(x)=(x/2)^l \sum_{k=0}^\infty (x/2)^{2k}/(k! (k+l)!)$
(restricted to a finite but large enough number of terms).

The liquid-gas coexistence curve can be found at $T < T_{\rm c}$
by applying the Maxwell construction to pressure (14). Alternatively,
we can introduce the chemical potential using the relation $\mu={\cal
F}+P/\rho$ and Eqs.~(8), (11), and (14). Then one obtains
\begin{equation}
{\cal \mu}={\cal \mu}_\varphi - k_{\rm B} T
\ln \int \exp(\beta {\bf h} \cdot {\bf s}) \, {\rm d} {\bf s} \, ,
\end{equation}
where
\begin{equation}
\int \exp(\beta {\bf h} \cdot {\bf s}) \, {\rm d} {\bf s} = \left\{
\begin{array}{lc}
2 \cosh\Big(\frac{H+a\rho m}{k_{\rm B} T}\Big)
\, , \ \ \ & n = 1 \, , \nonumber \\ [12pt]
2 \pi I_0\Big(\frac{H+a\rho m}{k_{\rm B} T}\Big)
\, , \ \ \ & n = 2 \, , \nonumber \\ [12pt]
\displaystyle
4 \pi \frac{\sinh\Big(\frac{H+a \rho m}{k_{\rm B} T}\Big)}
{\frac{H+a \rho m}{k_{\rm B} T}}
\, , \ \ \ & n = 3 \, , \nonumber \\ [12pt]
\end{array}
\right.
\end{equation}
and $\mu_\varphi$ is the chemical potential of the reference system
which should take its Carnahan-Starling form
\begin{equation}
\mu_\varphi=k_{\rm B} T \bigg[ \ln \rho +
\frac{\eta (8 - 9 \eta + 3 \eta^2)}{(1-\eta)^3} \bigg]
\end{equation}
to be self-consistent with Eq.~(15). Note that the chemical potential
and the pressure factor $Z_\varphi(\eta,T)=P_\varphi(\rho,T)/(\rho
k_{\rm B} T)$ are connected by the (exact) relation $\beta \mu_\varphi
(\eta,T)=\int_0^\eta [Z_\varphi(\eta',T)-1]/\eta' {\rm d} \eta'+
Z_\varphi-1+\ln(\Lambda \rho)$ (the term $k_{\rm B} T \ln \Lambda$,
with $\Lambda$ being the thermal de Broglie wavelength, has been
excluded from the right-hand side of Eq.~(23), since it depends only
on $T$ and is irrelevant for our consideration, see Eq.~(24) below).
The gas and liquid coexistence densities $\rho_{\rm G}(T)$ and
$\rho_{\rm L}(T)$ are then determined applying the well-known
mechanical and chemical equilibrium conditions
\begin{equation}
P(\rho_{\rm G},T)=P(\rho_{\rm L},T) \, , \ \ \ \
\mu(\rho_{\rm G},T)=\mu(\rho_{\rm L},T) \, .
\end{equation}

In the regime of large magnetic fields, we can solve the MES (13)
analytically taking into account the smallness of $\zeta=k_{\rm
B}T/(H+a \rho m) \ll 1$. This gives
\begin{equation}
m_{\zeta \ll 1} = \left\{
\begin{array}{lc}
1-2 \, {\rm e}^{-\frac{2 (H+a \rho)}{k_{\rm B}T}}
+{\cal O}\big({\rm e}^{-\frac{4}{\zeta^2}}\big)
\, , \ \ \ & n = 1 \, , \nonumber \\ [12pt]
1-\frac{1}{2} \frac{k_{\rm B}T}{H+a \rho} +
{\cal O}(\zeta^2)
\, , \ \ \ & n = 2 \, , \\ [15pt]
1-\frac{k_{\rm B}T}{H+a \rho} +
{\cal O}(\zeta^2)
\, , \ \ \ & n = 3 \, , \nonumber
\end{array}
\right.
\end{equation}
where the terms of the second and higher orders for ${\rm e}^{-2/\zeta}$
and $\zeta$ have been omitted, and the inequality ${\rm e}^{-2/\zeta} \ll
\zeta$ has been used. Substituting (25) into the PES (14), taking the
derivative of $P$ with respect to $\rho$, and solving the resulting
equation $(\partial P/\partial \rho)_{T,H}=0$ for $T$, we obtain the
critical temperature as a function of $H$. The result is
\begin{equation}
\frac{T_{\rm c}(H)\!-\!T_{\rm c\infty}}{T_{\rm c\infty}} = \left\{
\begin{array}{lc} 4 (W'_{\rm c\infty}-1) \,
{\rm e}^{-\frac{2 H}{k_{\rm B}T_{\rm c\infty}}}
\, , \ \ \ & n = 1 \, , \nonumber \\ [12pt]
\displaystyle
-\frac{k_{\rm B} T_{\rm c\infty}}{H}
\, , \ \ \ & n = 2 \, , \\ [15pt]
\displaystyle
-\frac{2 k_{\rm B} T_{\rm c\infty}}{H}
\, , \ \ \ & n = 3 \, , \nonumber
\end{array}
\right.
\end{equation}
where $T_{\rm c\infty}=\lim_{H \to \infty} T_{\rm c}(H)=
a \rho_{\rm c\infty}/(k_{\rm B} W'_{\rm c\infty})$ and
$\rho_{\rm c\infty}=\lim_{H \to \infty} \rho_{\rm c}(H)$
are the critical temperature and density in the infinite
magnetic field limit, $W'_{\rm c\infty}=\partial/\partial
\eta \, [ \eta (1+\eta+\eta^2-\eta^3)/(1-\eta)^3]\big|_{\eta=
\eta_{\rm c\infty}}$ with $\eta_{\rm c\infty}=\pi \rho_{\rm
c\infty} \sigma^3/6$, and it was assumed that the external
field is much larger than both the kinetic energy and the
internal magnetic field, i.e., $H \gg k_{\rm B} T_{\rm c\infty}$
and $H \gg a \rho_{\rm c\infty}$. From Eq.~(26) we conclude that
with increasing the external field, the critical temperature
approaches its limiting value from the top, when $n=1$, and from
the bottom, when $n=2$ or 3. Note that the factor $W'(\eta)-1>0$
is positive for all physical densities $\eta < 1$.
Moreover, for the Ising fluid model, the critical temperature
$T_{\rm c}(H)$ tends to $T_{\rm c\infty}$ exponentially with
increasing $H$. This is not the case for XY and Heisenberg fluids,
where $T_{\rm c}(H)$ reaches $T_{\rm c\infty}$ slower, according
to the inverse power law $H^{-1}$.

It is important to remark that in the saturation limit of infinite
magnetic field $H \to \infty$, all the spin fluid models considered
reduce to the same (nonmagnetic) fluid with the interparticle potential
$\phi(r)=\varphi(r)-J(r)$ consisting of the hard- or soft-core repulsion
part as well as the Yukawa-like attraction. The reason is that then the
spins align exactly along the field vector, so that the scalar product
${\bf s}_i \cdot {\bf s}_j$ will be equal to 1 (see Eq.~(1)) for any
pairs of particles. The term ${\bf H} \cdot \sum_i {\bf s}_i$ will
tends to a (infinite) constant and thus can be ignored in Eq.~(1)
(because we are entitled to accept a new level for counting the energy
of the system). The MF theory also leads to identical results for each
$n=1$, 2, and 3, when $H \to \infty$. Indeed, it follows from Eqs.~(13)
and (25) that the limit $\lim_{H \to \infty} m=1$ is independent of $n$.
The critical temperature and density at such a magnetic saturation, can
be found as usually, using the general relation (20) with putting $m=1$
in Eq.~(14). This yields $\rho^{\ast {\rm HS}}_{\rm c\infty} \approx
0.249$ and $T^{\ast \, {\rm HS}}_{\rm c\infty} \approx 2.264$ as well
as $\rho^{\ast \, {\rm SC}}_{\rm c\infty}\approx 0.262$ and $T^{\ast \,
{\rm SC}}_{\rm c\infty} \approx 2.380$ for the HS and SC versions,
respectively. Note also that in the limit $H \to \infty$, expressions
(21) and (22) for the chemical potential can be reduced (by extracting
an infinite constant depending only on $H$) to the form $\mu=
\mu_\varphi-a\rho$ that corresponds to a nonmagnetic system with
the potential $\phi$.

Simulations show (see the next section) that for the Yukawa-fluid
(YF) potential $\phi(r)=\varphi(r)-J(r)$ with the SC repulsion,
the critical liquid-gas temperature is equal to $T^{\ast}_{\rm YF}
\approx 2.680$. It is somewhat higher (within 12\%) than the temperature
$T^{\ast \, {\rm SC}}_{\rm c\infty} \approx 2.380$ obtained within the
SC version of the MF theory. It can be assumed that a significant part
of the above temperature discrepancy may come from the approximate form
used for the equation of state (see Eq.~(15), where $\eta=\xi \pi
\rho^\ast/6$) of the SC reference system. Therefore, it becomes quite
natural to introduce an adjustable SC (ASC) version of the MF theory,
where the second virial parameter $\xi(T)$ of the reference system
(Eq.~(17)) is replaced by its rescaled analog $\tilde \xi(T)=b\xi(T)$.
The constant $b$ can then be determined by requiring the critical
temperature $T^{\ast}_{\rm YF}(b)$ coincides with the exact result.
This leads to $b \approx 0.902$ and corresponds to a slight decrease
(by a factor of $b^{1/3} \approx 0.966$) of the effective diameter
when calculating the pressure according to Eq.~(15) with $\eta=
\tilde \xi(T) \pi \rho^\ast/6$.

\vspace{6pt}

\section{Computer simulation and theory calculations}

\subsection{Simulation procedures}

Two kinds of MC simulations have been carried out to investigate
the critical behavior. First, in order to determine the liquid-gas
coexistence properties we have applied the Gibbs ensemble MC (GEMC)
approach \cite{Panagio} within an advanced biasing scheme \cite{Lom}
for handling spin degrees of freedom. The GEMC simulations were
performed for two system sizes, namely, with $N=500$ and $1000$
particles (or only with $N=500$ or $1000$ for some temperatures
and nonzero values of $H$ to save computer time). In each case,
the particles were distributed over two boxes with volumes which
fluctuated under the constraint of fixed total volume. The GEMC
configurations were generated in cycles, where one cycle consisted
of either (i) $N$ trial displacements and spin reorientations of
particles chosen at random; or (ii) one attempted volume rearrangement
of the boxes; or (iii) $N_{\rm ex}$ attempts to exchange the particles
between the simulation boxes. The type of each cycle was selected at
random and with equal probability among the above three possible steps.

The bias has been used during the spin reorientation and exchange
(insertion) steps, so that a new attempted direction for vector ${\bf
s}_i$ was generated with a probability that favors orientations parallel
to the local magnetic field. This bias was taken into account when
considering the acceptance probabilities corresponding to steps (i) and
(iii) as well as when calculating the chemical potential. The analytical
expressions for such probabilities have been obtained by modifying the
hard-sphere acceptance rules \cite{Lom} to the case when the soft-
(instead of hard-) core nonmagnetic repulsion potential is present.
The acceptance ratios for the particle moves and volume changes were
adjusted to lie in a range of 30 -- 60\%. The value for $N_{\rm ex}$
was chosen to yield a success rate of particle transfers of 0.1 -- 3\%,
depending on the temperature. The chemical potential was calculated
during the exchange (insertion) step using the generalized Widom's
method \cite{Frenkel}, while the pressure was obtained employing the
virial theorem.

Secondly, the Binder crossing technique \cite{Mryglod,Binder} has been
utilized to study the magnetic phase transitions (in the absence of an
external magnetic field). Here, the usual canonical MC simulations have
been performed for system sizes of $N_1=250$ and $N_2=500$ particles at
several particle number densities $\rho$ and temperatures $T$. At each
value of $\rho$, the Binder parameter $B=1-M_4/(3 M_2^2)$ was plotted
as a function of $T$ for the two system sizes (this parameter represents
a fourth-order cumulant with $M_l=\langle s^l \rangle$ and $s=|\sum_i
{\bf s}_i|$). The critical temperature $T_\lambda(\rho)$ of the
para-ferro phase transition was then obtained from the position
of the intersection point of curves $B_{N_1}(T)$ and $B_{N_2}(T)$,
i.e., from the condition $B_{N_1}(T_\lambda)=B_{N_2}(T_\lambda)$. As
in the case of GEMC simulations, the orientational biasing technique
\cite{Lom} was used to improve the convergence of the canonical
calculations.

Within both the GEMC and canonical simulations, the Yukawa function was
truncated at half of the box edge. In order to reduce the finite-size
effects, a long range correction was taken into account (within periodic
boundary conditions) by including an additional term to the potential
energy of the system. This term was derived \cite{Lom} by integrating
the Yukawa function beyond the cutoff radius and assuming that the binary
distribution function is equal to unity in this region. For sufficiently
large system sizes (as in our case) such an assumption should lead to
virtually exact results. No additional finite size scaling corrections
were applied in determining the critical temperatures, assuming that
they are small on the scale of phenomena considered. The number of MC
cycles used to achieve an equilibrium state varied from $2 \times 10^4$
to $5 \times 10^5$ depending on the system and external field value.
Note that in the GEMC simulations, the pressure and chemical potential
should be the same (within statistical noise) at equilibrium in the
two (gas and liquid) boxes. After achieving the equilibrium, the
investigated quantities were measured by averaging over $2 \times
10^5$ to $2.5 \times 10^6$ cycles (depending on the model, ensemble,
thermodynamic point, and field strength). The statistical
uncertainties have been estimated using the block averages
method \cite{Frenkel}.

The critical temperature $T_{\rm c}$ and density $\rho_{\rm c}$ have been
evaluated at $H \ne 0$ by fitting the discrete set of GEMC data near the
criticality to the dependence $\rho_\pm=\rho_{\rm c}+C_1(1-T/T_{\rm c})
\pm C_2(1-T/T_{\rm c})^\beta$ (the constants $C_1$ and $C_2$ should
minimize the deviations between $\rho_\pm$ and simulation values). Here
the law of rectilinear diameters $(\rho_{\rm L}+\rho_{\rm G})/2=\rho_{\rm
c}+C_1(1-T/T_{\rm c})$, the power law behavior $\rho_{\rm L}-\rho_{\rm
G}=2C_2(1-T/T_{\rm c})^\beta$, and the critical exponent $\beta=0.32$
have been assumed as in Ref.~\cite{Kotelyanskii} (where results for a
Lennard-Jones fluid were analyzed). Note that for $H \ne 0$ the magnetic
para-ferro phase transition disappears and the liquid-gas coexistence
curves behave like those of nonmagnetic fluids with the effective
attraction potential $J_{\rm eff}(r)=-\langle {\bf s}_1 \cdot {\bf s}_2
\rangle_H J(r)$. At $H=0$, where both the liquid-gas and magnetic
para-ferro transitions exist simultaneously, the application of the
above fitting should be performed with care because the rectilinear
and power laws will not work properly at temperatures close enough
to the temperature $T_\lambda$ of the magnetic transition.

\subsection{Results and discussion}

\subsubsection{Ising fluid}

Examples of the liquid-gas coexistence obtained in the GEMC simulations
for the soft-core Ising fluid at various values, $H^\ast=H/\epsilon = 0$,
0.1, 0.5, 1, 5, and $\infty$, of the external field are shown in Fig.~1.
As can be seen clearly, the critical temperature $T_{\rm c}$ goes down
monotonously with rising the field strength and rapidly tends to its
minimal value in the infinite field limit. For instance, already at
$H^\ast=5$, the gas and liquid binodal branches are practically
indistinguishable from those corresponding to the case $H \to \infty$.
In addition, with increasing $H$ the shape of the coexistence curves
becomes wider near a critical point. At $H=0$, all the simulated points
belonging to the gas phase lie on a curve which is very close in form to
a straight line left to the magnetic para-ferro transition Curie line.
The intersection of these lines defines a critical end point which
coincides with the liquid-gas critical point expected for a tricritical
point, i.e., $T_{\rm ce}=T_{\rm c}=T_{\rm t}$. In other words, the Ising
fluid exhibits a tricritical behavior. The shape of the Ising binodals
has been verified by the multiple-histogram reweighting (MHR) method
\cite{Ferrenberg} and agreement with the GEMC results was observed.

\vspace{16pt}

\begin{figure}[htbp]
\epsfxsize=84mm
\centerline{\epsffile{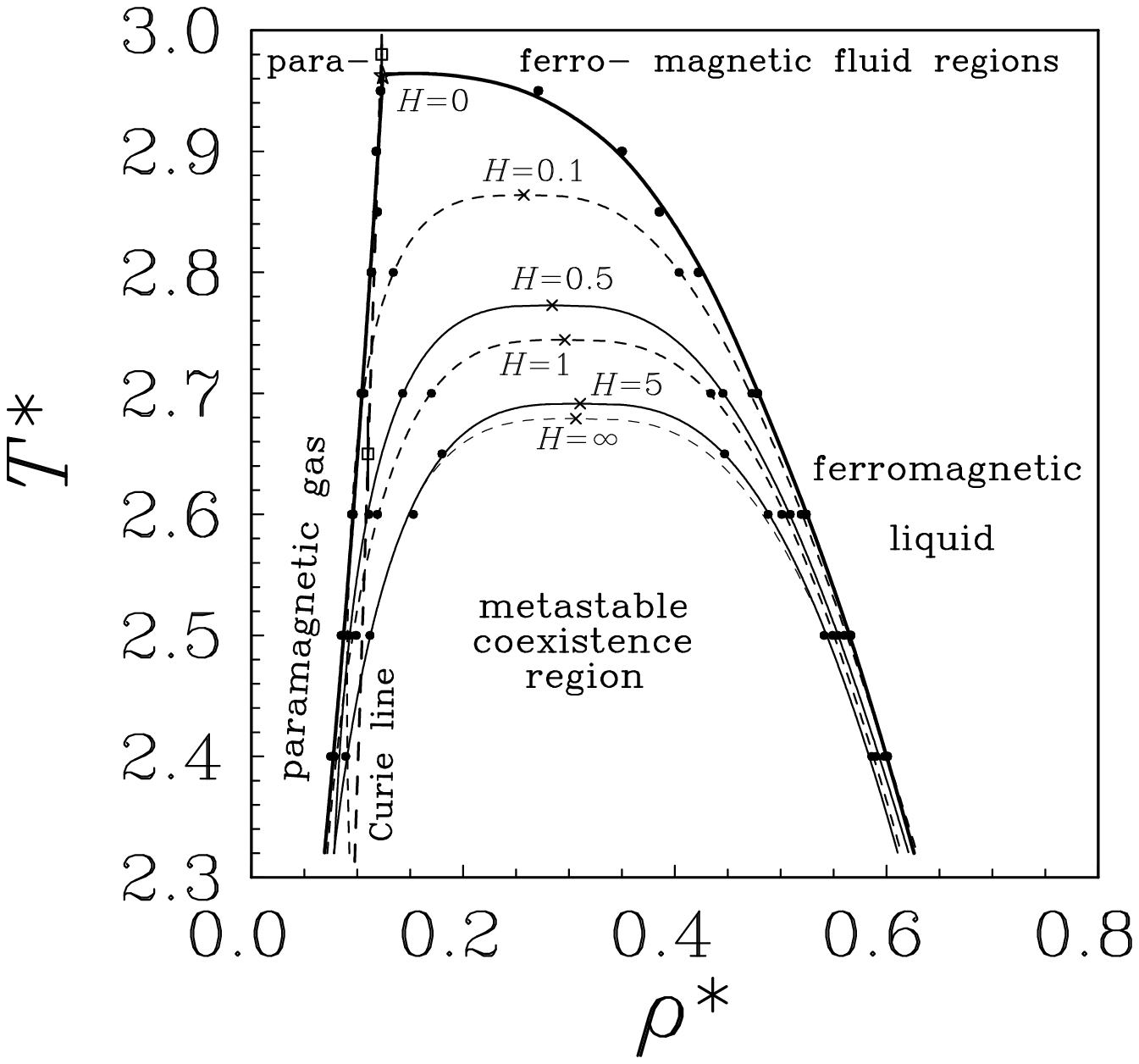}}

\vspace{6pt}

{\small FIG.~1. The liquid-gas coexistence curves obtained from GEMC
simulation data (circles) for a soft-core Ising fluid at various
values of the external magnetic field $H$. The para-ferro phase
transition (at $H=0$) is plotted by the dashed (Curie) line passing
via open squares (obtained in canonical MC simulations). The
critical points are shown as crosses, whereas the tricritical
point is presented by the star. The values of $H$ are given
in the dimensionless form $H^\ast=H/\epsilon$.}
\end{figure}

\vspace{9pt}

In making the above conclusion on the phase diagram topology at $H=0$
we cannot be absolutely sure, since the GEMC simulations (as well as
the MHR technique) do not provide us with precise enough data for
temperatures which are very close to the critical region. Near this
region, the particle fluctuations become too large, so that the two
GEMC boxes (which consist of finite numbers of particles) can switch
their identity many times during the simulations. This prevents one
from obtaining good mean density values of the system in gas and liquid
phases at $T \sim T_{\rm c}$ (although the density distributions over
the boxes can still indicate the existence of two phases). On the other
hand, having a discrete set of GEMC coexistence points lying relatively
far from the criticality region, it is impossible to apply the fitting
procedure (see at the end of the preceding subsection) in the vicinity
of the expected tricritical point, where $T \sim T_\lambda(\rho_{\rm
c}) = T_{\rm c}$, because then the liquid and magnetic transitions
are coupled. The question concerning the topology of critical points
at $H=0$ goes beyond the scope of the present paper and requires
additional investigations by more sophisticated simulation techniques
(such as finite-size scaling \cite{Fisher}, for example).

The primary goal of this work is to study the liquid-gas phase
transition in the presence ($H \ne 0$) of an external field with
focusing on the calculations of $T_{\rm c}$ as a function of $H$.
The results of these calculations obtained in the case of the Ising
fluid within HS, SC, and ASC versions of the MF theory are shown
in Fig.~2 together with the simulation data. Here, a more extended
set of external field values has been used. It can be seen that the
HSMF and SCMF schemes are able to predict qualitatively the monotonic
decrease of the critical temperature $T_{\rm c}$ with rising $H$. The
relative deviations between the simulation data and SCMF predictions
for $T_{\rm c}$ are of order \mbox{10 -- 15\%}, i.e., they are not so
large in view of assumptions made within the MF approach. When the
ASC version is used, the theoretical and simulation data appear to
be practically indistinguishable and the disagreements do not exceed
the GEMC statistical noise. At weak external fields ($H^\ast \lesssim
0.5$), the function $T_{\rm c}(H)$ behaves as $T_{\rm c}(0)-cH^{2/5}$
(see Fig.~2, where the latter dependence is plotted at $c=0.258$
for the ASC case). This confirms the MF prediction $\lim_{H \to 0}
[T_{\rm c}(H)-T_{\rm c}(0)] \propto - H^{2/5}$, obtained previously
in Refs.~\cite{Schinagl,Schinfol}. In the strong field regime ($H^\ast
\gtrsim 2$), the critical temperature decreases exponentially,
according to the analytical formula derived in sect.~II.C (see
the first line of Eq.~(26)).

\vspace{18pt}

\begin{figure}[htbp]
\epsfxsize=84mm
\centerline{\epsffile{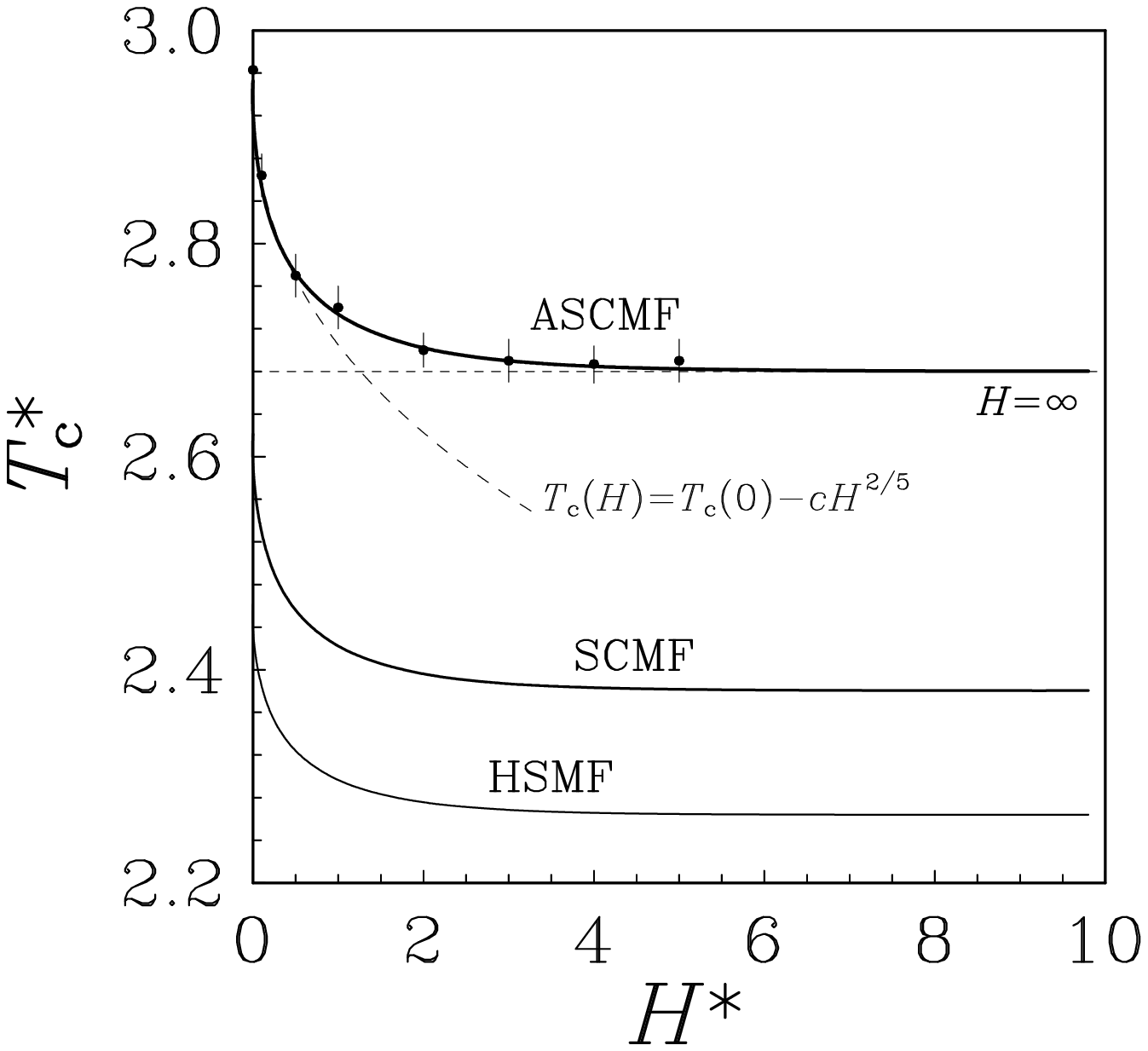}}

\vspace{8pt}

{\small FIG.~2. The critical temperature $T_{\rm c}$ as a function of
the external magnetic field $H$ obtained for the soft-core Ising fluid
from GEMC simulations (circles) in comparison with the results of
the HS, SC, and ASC versions of the MF theory (solid curves). The
critical level corresponding to the infinite value of the external
field is plotted by the horizontal dashed line. Vertical bars
indicate statistical uncertainties.}
\end{figure}

\vspace{9pt}

The Ising gas-liquid coexistence curves evaluated within the ASC
version of the MF theory as well as the MF para-ferro transition
line are presented in Fig.~3. The deviations in the theoretical and
simulation binodals (please compare Figs.~1 and 3) are larger than
the discrepancy in the case of function $T_{\rm c}(H)$ (see Fig.~2)
and they cannot be reduced to zero even within the \mbox{ASCMF}. For
instance, the theory somewhat overestimates the gas phase densities
and underestimates the densities in the liquid phase. As a result,
the shape of the binodals appears to be narrower with respect to that
of the simulation coexistence curves. A more accurate theory should
be applied to describe quantitatively the liquid-gas phase diagrams
in the whole density, temperature, and magnetic field ranges.

\vspace{17pt}

\begin{figure}[htbp]
\epsfxsize=84mm
\centerline{\epsffile{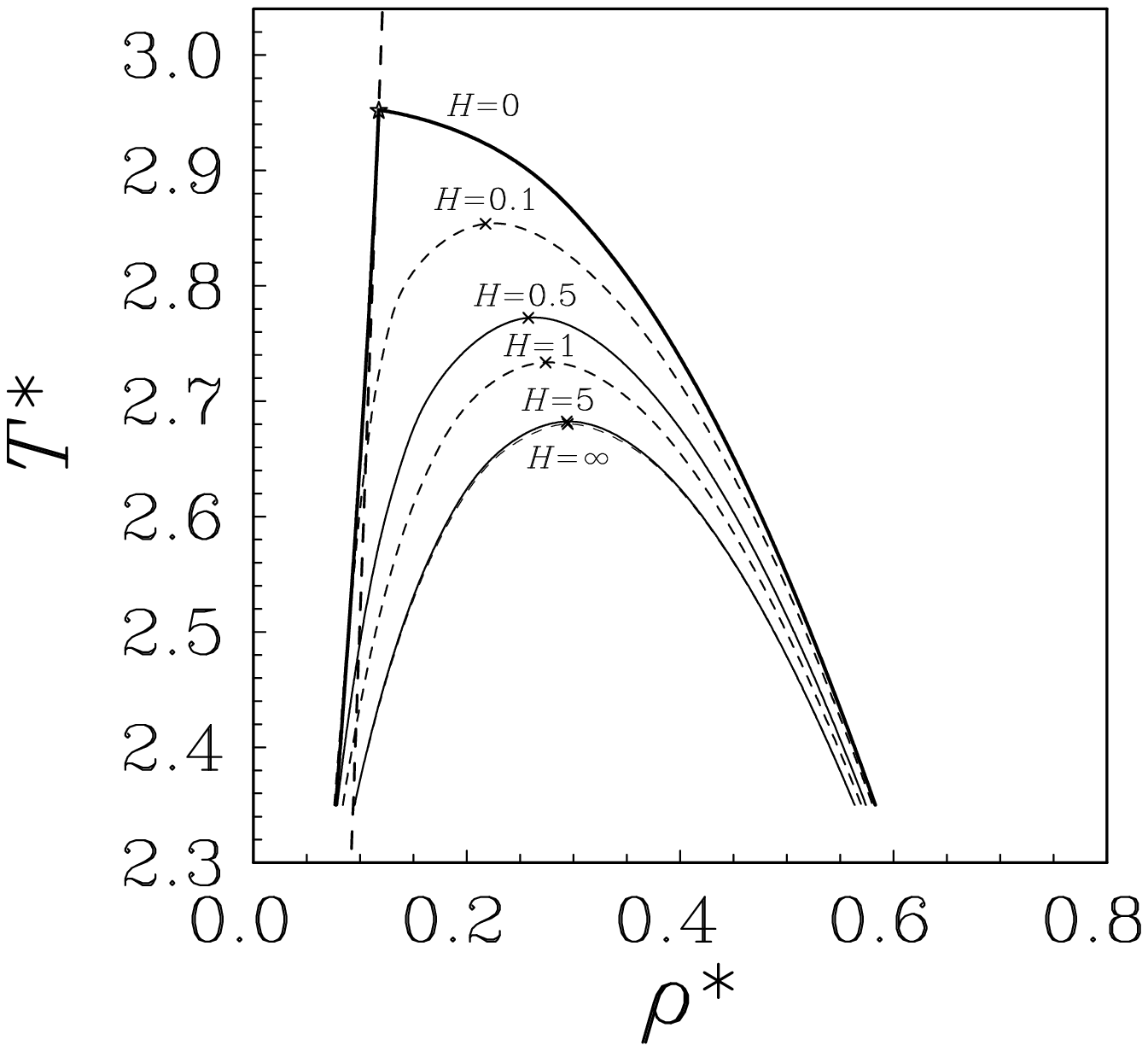}}

\vspace{6pt}

{\small FIG.~3. The gas-liquid coexistence of the Ising fluid obtained
within the ASC version of the MF theory. The MF para-ferro transition
is plotted by the dashed line. For other notations see the caption to
Fig.~1.}
\end{figure}

\vspace{7pt}

\subsubsection{XY-spin fluid model}

As was pointed out in the introduction, until now no computer experiment
and theoretical investigations on the liquid-gas coexistence have been
performed for the planar XY spin fluid model. In this respect it should
be mentioned that in order to obtain the coexistence curves within the
MF theory, one has to solve numerically Eq.~(24). At the same time, the
HS, SC, and ASC functions $T_{\rm c}(H)$ are obtained by finding numerical
solutions to Eq.~(20) at given values of $H$. Taking into account that
the magnetic equation of state (13) must be handled numerically too, the
calculation of $\rho_{\rm L,G}(T)$ and $T_{\rm c}(H)$ presents, in fact,
a rather complicated technical problem. This is especially true in the
case of XY system ($n=2$) and SC/ASC versions of the MFT, where the
integration in Eq.~(17) as well as the computation of Bessel's functions
are required additionally. Similar difficulties arise in MC simulations
of XY fluids when applying the orientational biasing technique \cite{Lom}.
The reason is that then the trial orientation vectors should be generated
with distributions, which cannot be presented (contrary to the cases $n=1$
and $n=3$) by simple algebraic expressions, and the use of time-consuming
Bessel-like functions is required. Nevertheless, developing an efficient
algorithm has allowed us to overcome the technical difficulties and
calculate the coexistence curves for all the models, including the XY.

\vspace{18pt}

\begin{figure}[htbp]
\epsfxsize=84mm
\centerline{\epsffile{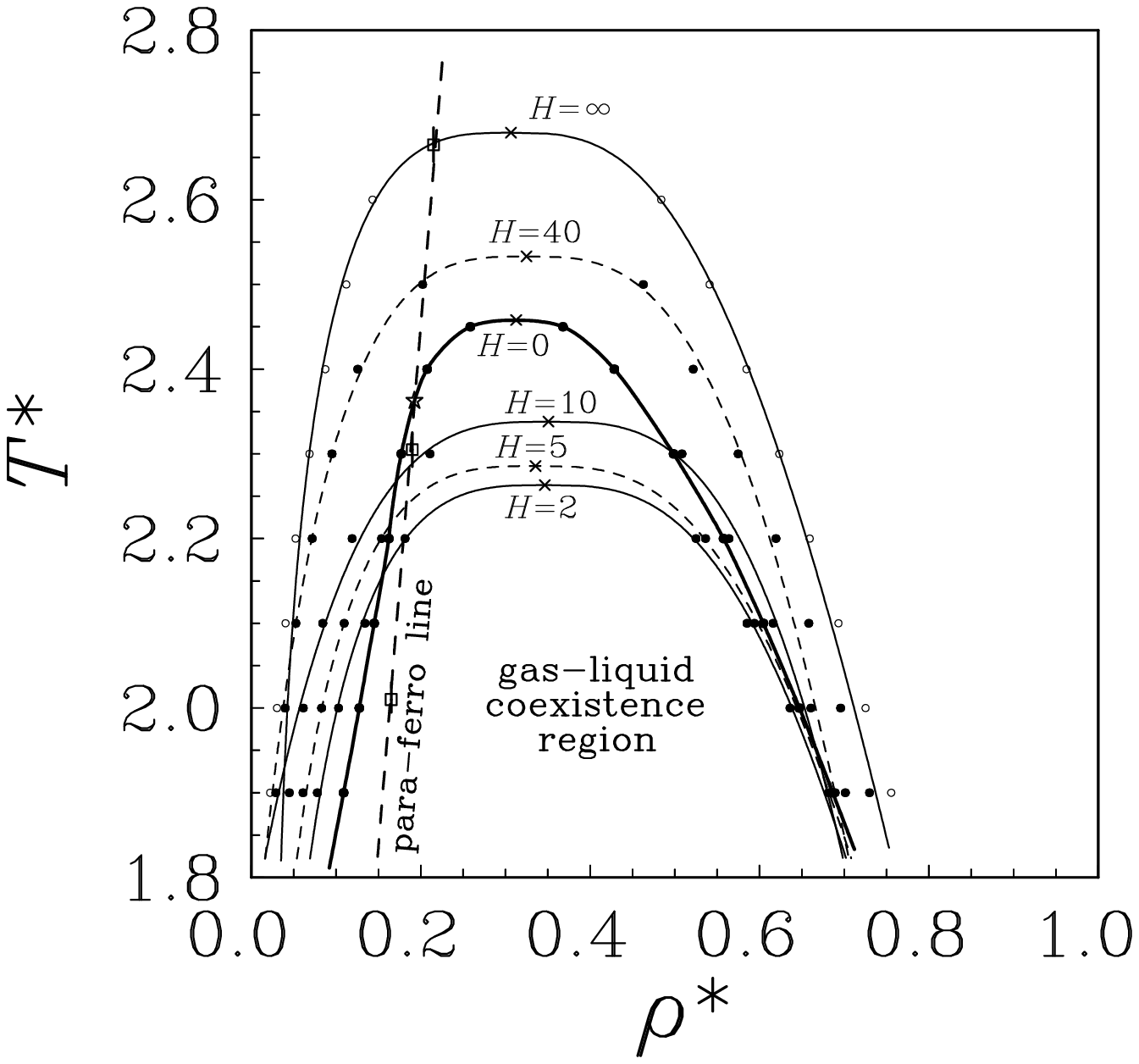}}

\vspace{6pt}

{\small FIG.~4. The gas-liquid (curves) and para-ferro (dashed line)
phase coexistences obtained from GE (at different values of an external
field $H$) and canonical (at $H=0$) MC simulations, respectively, for
the XY-spin fluid model. The critical end point is shown by the star.
Other notations are the same as in Fig.~1.}
\end{figure}

\vspace{9pt}

The XY gas-liquid binodals evaluated in the GEMC simulations are presented
in Fig.~4 for the set $H^\ast=0$, 2, 5, 10, 40, and $\infty$ of external
field values. As can be clearly seen, here the topology of phase diagrams
differs in several aspects from that of the Ising fluid. First, at $H=0$
the para-ferro magnetic line (which is included in Fig.~4 as well)
intersects the gas branch of the binodal at a critical end point, $T_{\rm
ce}$, which does not coincide with the critical point $T_{\rm c}$, i.e.,
$|T_{\rm c}-T_{\rm ce}| \ne 0$. So that, contrary to the Ising fluid,
where the gas can only be in a paramagnetic state and the liquid be only
ferromagnetic, the XY system exhibits a richer pattern. Here, the gas can
be either in paramagnetic (when $T < T_{\rm ce}$) or ferromagnetic (when
$T_{\rm ce} < T < T_{\rm c}$) states, and thus the liquid-gas transition
can take place with keeping the ferromagnetic ordering. Secondly, with
increasing $H$, the critical temperature starts to decrease rapidly
reaching its minimal value at $H^\ast \sim 2$ and further begins to
increase much slower tending to the infinite-field limit $T_{\rm
c\infty}$. The shape of the binodals near $T_{\rm c}(H)$ also
demonstrates a nonmonotonic behavior, namely, it first becomes
wider with rising $H$ and then begins to narrow, but the most
sharp shape remains at $H=0$.

\vspace{18pt}

\begin{figure}[htbp]
\epsfxsize=84mm
\centerline{\epsffile{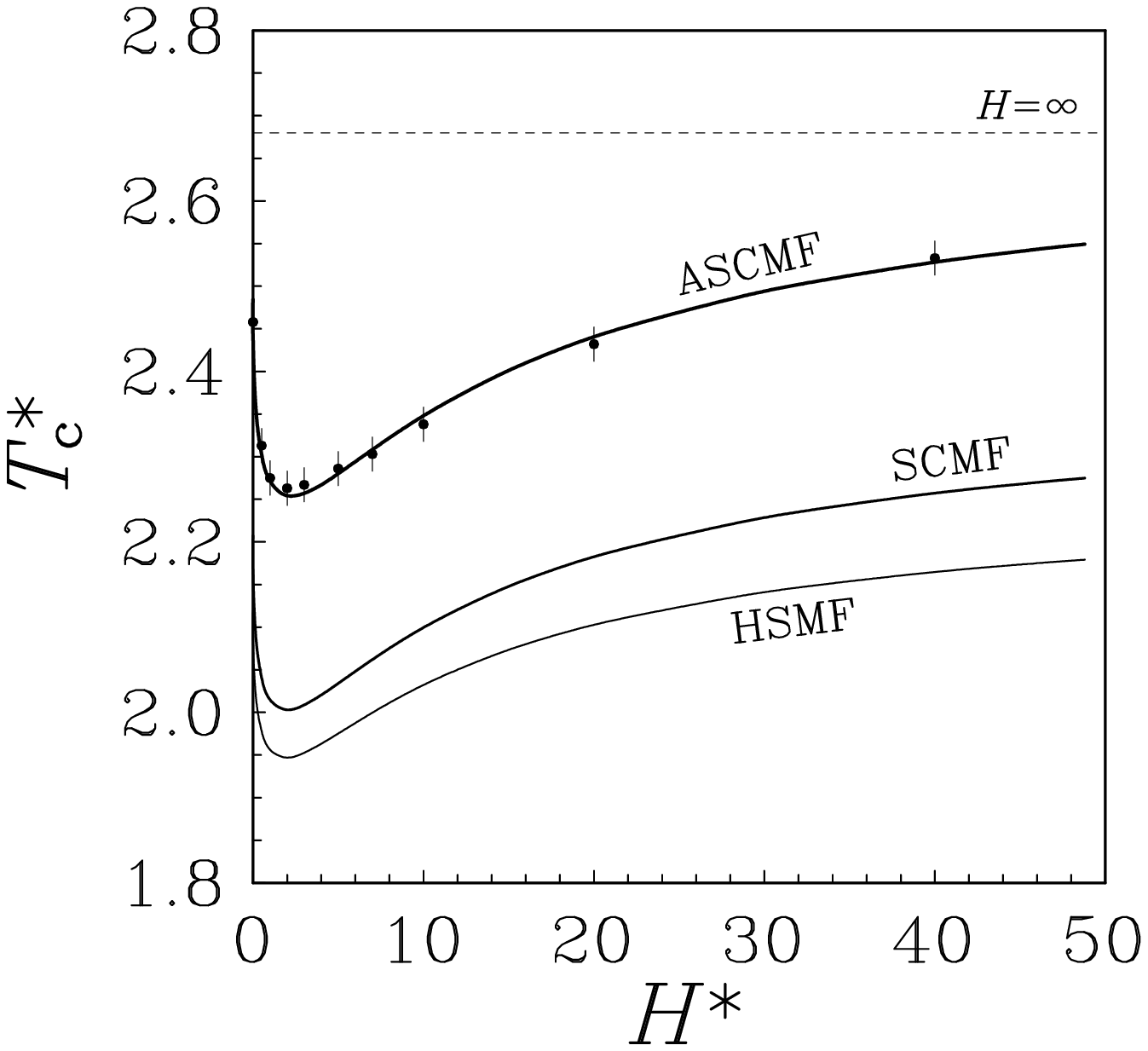}}

\vspace{6pt}

{\small FIG.~5. The critical temperature as a function of the external
field obtained for the XY-spin fluid model from GEMC simulations
(circles) in comparison with the results of the HS, SC, and ASC
versions of the MF theory (solid curves). Other notations are the
same as in Fig.~2.}
\end{figure}

\vspace{12pt}

The nonmonotonic dependence of $T_{\rm c}$ on the strength $H$ of the
external field observed from the simulations for the XY fluid is shown
in more detail in Fig.~5. The corresponding results of the HS, SC, and
ASC MF approaches are included there too. Again we can see that the HS
and SC versions lead to qualitatively correct results and the deviations
from the simulation data do not exceed about 15\%. In addition, all the
approaches predict a minimum of $T_{\rm c}(H)$ at nearly the same
external field $H^\ast \approx 2$. Moreover, the ASC approach provides
us with virtually exact results for $T_{\rm c}(H)$ at any value of $H$.
For instance, the theoretical discrepancy is less than the level of
GEMC uncertainties. However, at $H \to 0$ one has to be careful in
application of the MF $H^{2/5}$-dependence (see subsection III.B.{\em
1}) to interpretation of the function $T_{\rm c}(H)$ build on simulation
data. Here, a crossover to an analytic behavior with $\lim_{H \to 0}
\partial T_{\rm c}/\partial H =0$ is expected because of the lack
of a tricritical point ($|T_{\rm c}-T_{\rm ce}| \ne 0$). For large
fields ($H^\ast \gtrsim 50$), the critical temperature tends to the
infinite-field value $T_{\rm c\infty}$ as $T_{\rm c}(H)-T_{\rm c\infty}
\propto -1/H$, which is consistent with MF prediction (26) at $n=2$.
The saturation regime, when $T_{\rm c}(H)$ is almost equal to $T_{\rm
c\infty}$, can be achieved here by very strong fields of order $H^\ast
\gtrsim 500$ (for the Ising fluid the saturation level is much lower,
$H^\ast \gtrsim 5$, see Fig.~2).

\vspace{18pt}

\begin{figure}[htbp]
\epsfxsize=84mm
\centerline{\epsffile{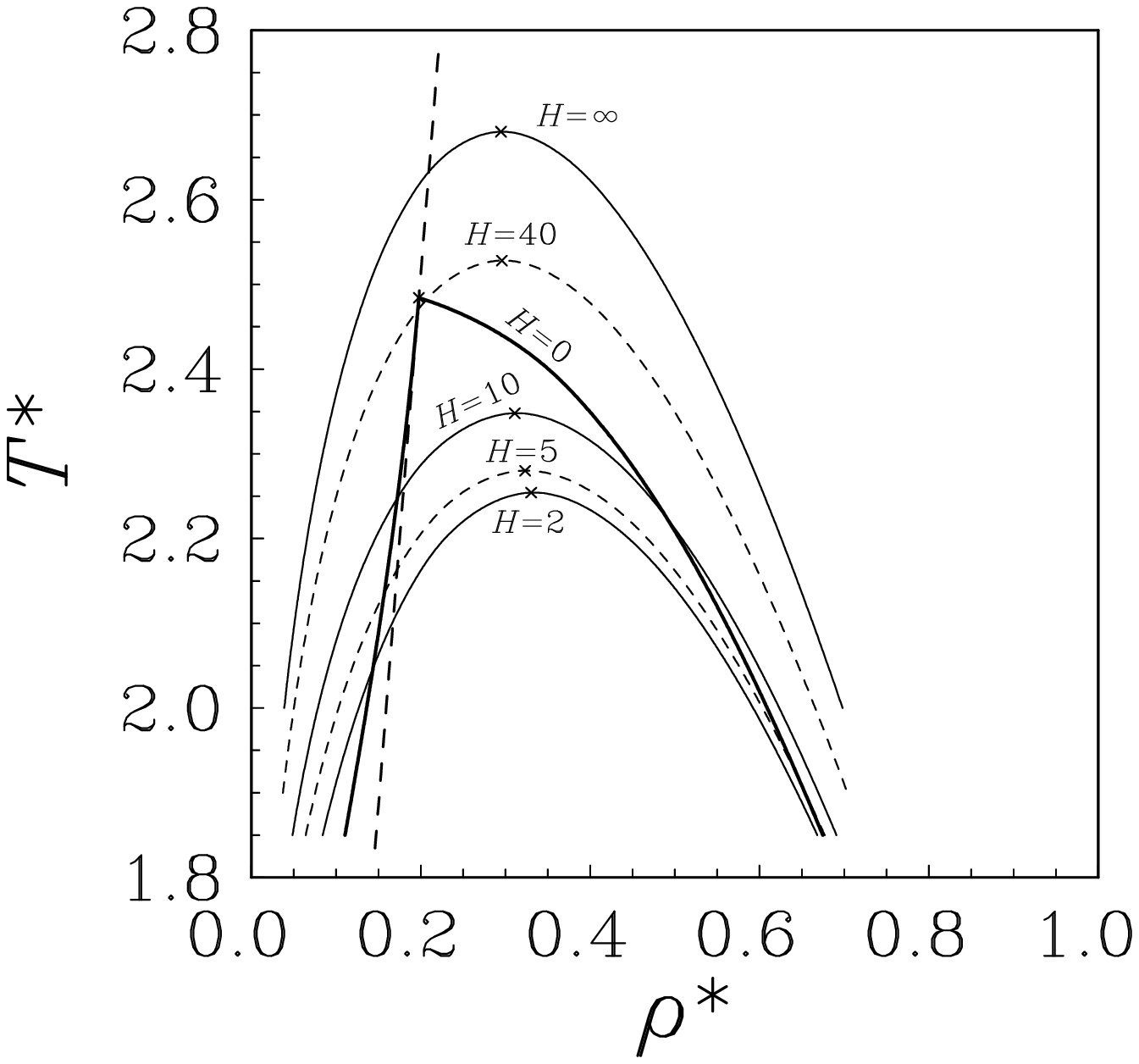}}

\vspace{6pt}

{\small FIG.~6. The gas-liquid and para-ferro coexistences of the XY-spin
fluid observed within the ASC version of the MF theory. Notations are
similar to those of Fig.~4.}
\end{figure}

\vspace{12pt}

The theoretical liquid-gas coexistence curves and para-ferro transition
line of the XY fluid are shown in Fig.~6. Comparing these results with
the corresponding GEMC data (see Fig.~4) we see that, as in the case of
the Ising system, the ASCMF predictions of $\rho_{\rm G}$ and $\rho_{\rm
L}$ are not as perfect as those of the field dependency of $T_{\rm c}$.
Again, at $H \ne 0$ the shape of the binodals becomes narrower because
of the overestimation and underestimation of gas and liquid densities,
respectively. For $H=0$, the ASCMF approach does not predict the
existence of critical and critical end points found in the simulations,
but leads instead to a tricritical behavior.

\subsubsection{Heisenberg fluid}

The liquid-gas coexistence curves and the Curie line obtained for the
Heisenberg fluid within the simulations and ASCMF theory are shown in
Figs.~7 and 8, respectively. As can be seen, the Heisenberg system
exhibits a topology of phase diagrams similar to the XY fluid. In
particular, according to the simulation results for $H=0$, the Curie
line ends at a critical end point on the gas side of the binodal, so
that $T_{\rm ce} < T_{\rm c}$, where $T_{\rm c}$ should be referred
to the liquid-gas critical point located in the ferromagnetic phase
(see Fig.~7). It is worth mentioning that some evidence of the lack
of a tricritical point in the Heisenberg system has been provided
by GEMC simulations earlier \cite{Lom}. Later on, it was stated
\cite{Weis} that owing to finite size effects it is very difficult
to come definitely to one of the two possible scenarios: whether the
Curie line ends at a critical end point (as suggested by the simulations,
see Fig.~7), or at a tricritical point (as suggested by the MF theory,
see Fig.~8). Relatively recently, the existence of a critical end point
for the Heisenberg fluid has been observed within the integral equation
theory \cite{Sokolovskii}. However, the difference $|T_{\rm ce} - T_{\rm
c}|$ reported there was very small and amounted to about 0.1\%, which
can be comparable with numerical uncertainties. On the other hand, this
difference is much smaller than the discrepancy introduced by the mean
spherical approximation, used in Ref.~\cite{Sokolovskii} for the closure
relation to the Ornstein-Zernike equation.

\vspace{12pt}

\begin{figure}[htbp]
\epsfxsize=84mm
\centerline{\epsffile{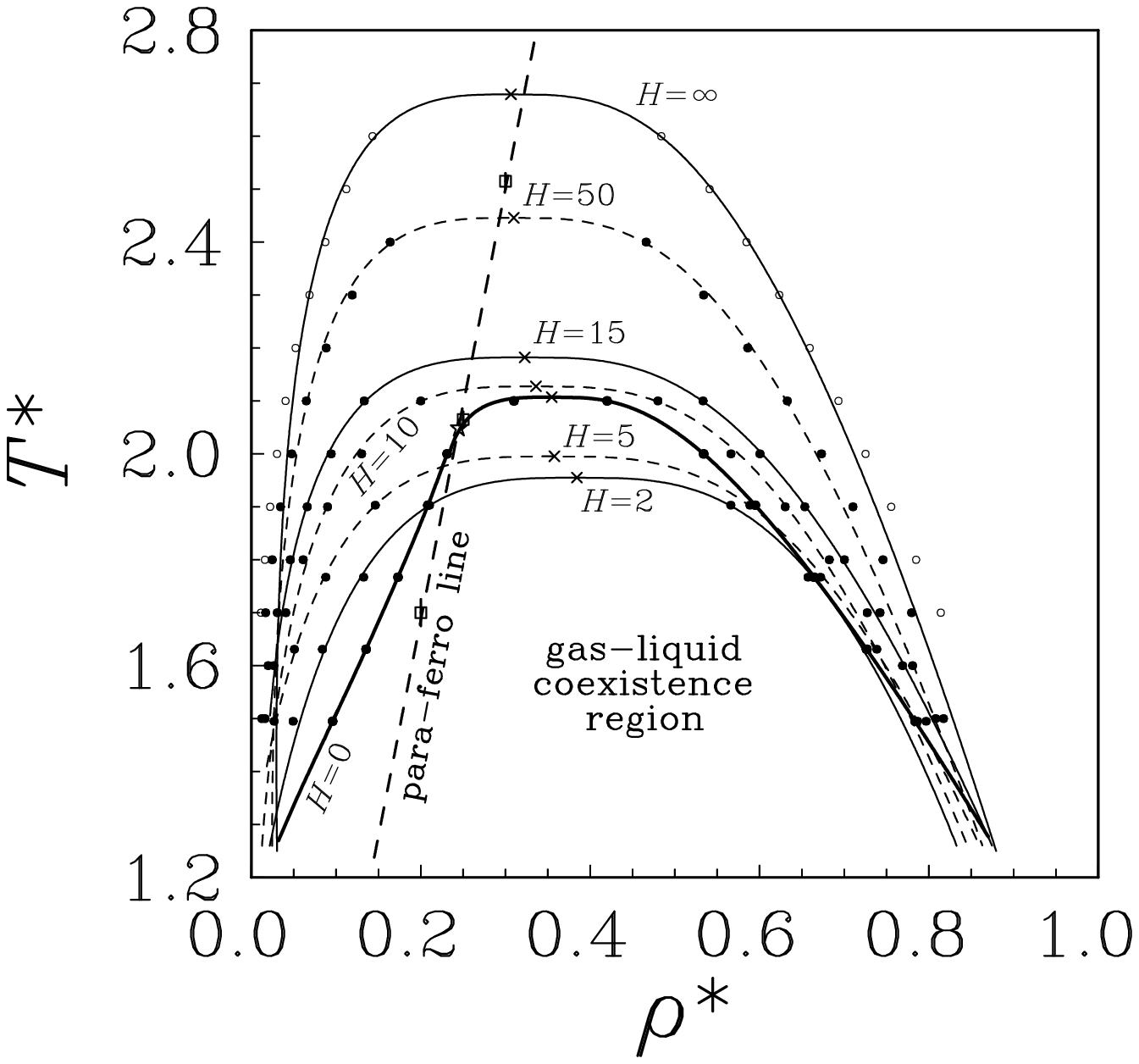}}

\vspace{6pt}

{\small FIG.~7. The gas-liquid (curves) and para-ferro (dashed line)
phase coexistence obtained from GE (at different values of an external
field $H$) and canonical (at $H=0$) MC simulations, respectively, for
the soft-core Heisenberg fluid. Other notations are the same as in
Fig.~4.}
\end{figure}

\vspace{12pt}

In the presence of an external field, the change in shape of the
liquid-gas binodal and the change of the critical temperature are
nonmonotonic for the Heisenberg fluid like for the XY system. This
can be seen from Fig.~7 for the set $H^\ast=0$, 2, 5, 10, 15, 50, and
$\infty$, as well as from Figs.~9 and 10, where the values of $T_{\rm
c}(H)$ are presented for a more complete set of $H$ and in more detail
at small $H$, respectively. Again, the theory is able to qualitatively
describe the liquid-gas coexistence properties (see Fig.~8), whereas
the function $T_{\rm c}(H)$ can be calculated within the ASCMF approach
quantitatively. The HS and SC versions of the MF theory also reproduce
well the critical temperature, but they underestimate $T_{\rm c}(H)$
to within an order of 10\%. The minimum of $T_{\rm c}(H)$ at $H^\ast
\approx 2$ can be predicted by either version of the MFT. It should
be pointed out also that the presence of SC repulsion leads to an
increase of the gas-liquid critical temperature with respect to that
of the HS potential (the SCMF curve lies above the HSMF at all values
of $H$, see Figs.~9 and 10). The same conclusion is valid for the
Ising and XY models (see Figs.~2 and 5). The asymptotic behavior of
$T_{\rm c}(H)$ at $H^\ast \gg 1$ looks like $T_{\rm c}(H)- T_{\rm
c\infty} \propto -1/H$. It is identical in form to that of the XY
system, but differs in the value of the coefficient of the
proportionality to $1/H$. Because this coefficient is twice
as large at $n=3$ (see Eq.~(26)), the saturation regime will
begin here at stronger fields ($H^\ast \gtrsim 1000$).

\vspace{12pt}

\begin{figure}[htbp]
\epsfxsize=84mm
\centerline{\epsffile{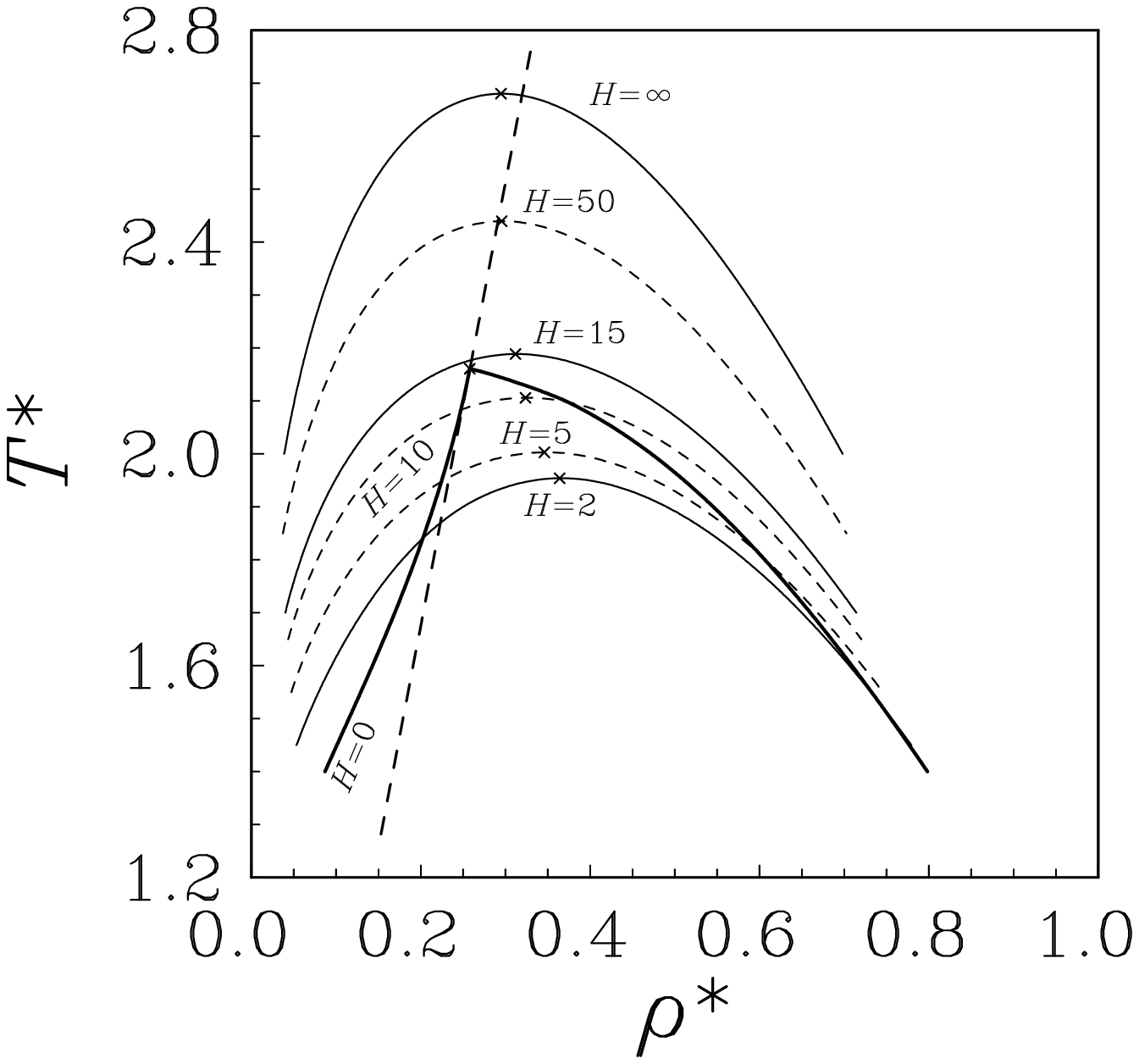}}

\vspace{6pt}

{\small FIG.~8. The gas-liquid and para-ferro coexistences of the
Heisenberg fluid obtained within the ASC version of the MF theory.
Notations are the same as in Fig.~6.}
\end{figure}

\vspace{12pt}

Note also that in the limit $H \to 0$, the MF function $T_{\rm c}(H)$
behaves like $\sim -H^{2/5}$ (with $\partial T_{\rm c}/\partial H \sim
-H^{-3/5} \to -\infty$) independently of $n$ (see section IV). However,
as in the case of XY fluid, the function $T_{\rm c}(H)$ corresponding
to simulation data for the Heisenberg model (see Figs.~9 and 10) should
exhibit an analytic bahavior at $H \to 0$ with $\lim_{H \to 0} \partial
T_{\rm c}/\partial H=0,$ because of the existence of a critical end
point ($|T_{\rm c}-T_{\rm ce}| \ne 0$).

\vspace{12pt}

\begin{figure}[htbp]
\epsfxsize=84mm
\centerline{\epsffile{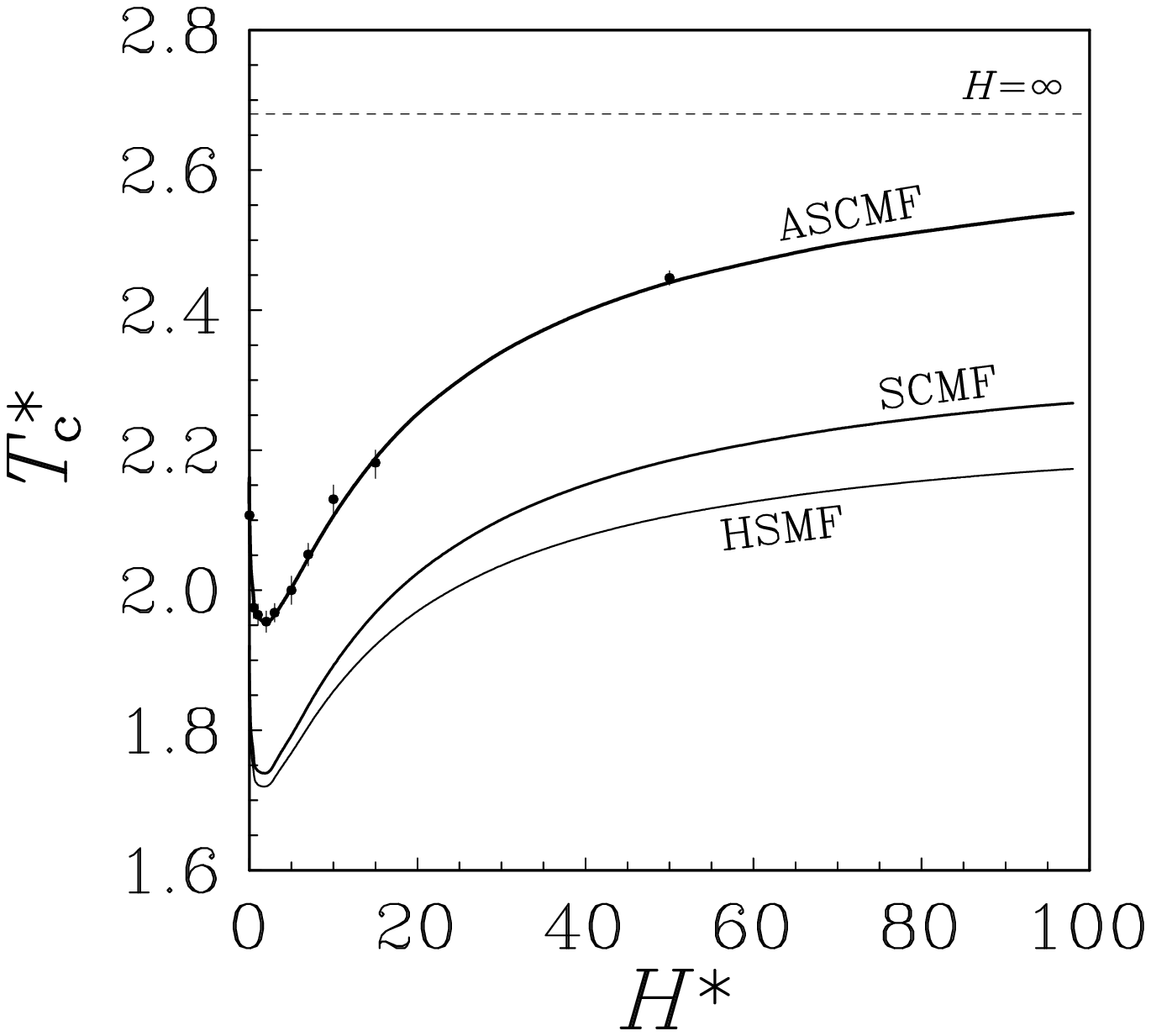}}

\vspace{6pt}

{\small FIG.~9. The critical temperature $T_{\rm c}$ as a function
of the external field $H$ obtained for the Heisenberg fluid from GEMC
simulations (circles) in comparison with the results of the HS, SC,
and ASC versions of the MF theory (solid curves). Other notations
are the same as in Fig.~5.}
\end{figure}

\vspace{12pt}

\begin{figure}[htbp]
\epsfxsize=84mm
\centerline{\epsffile{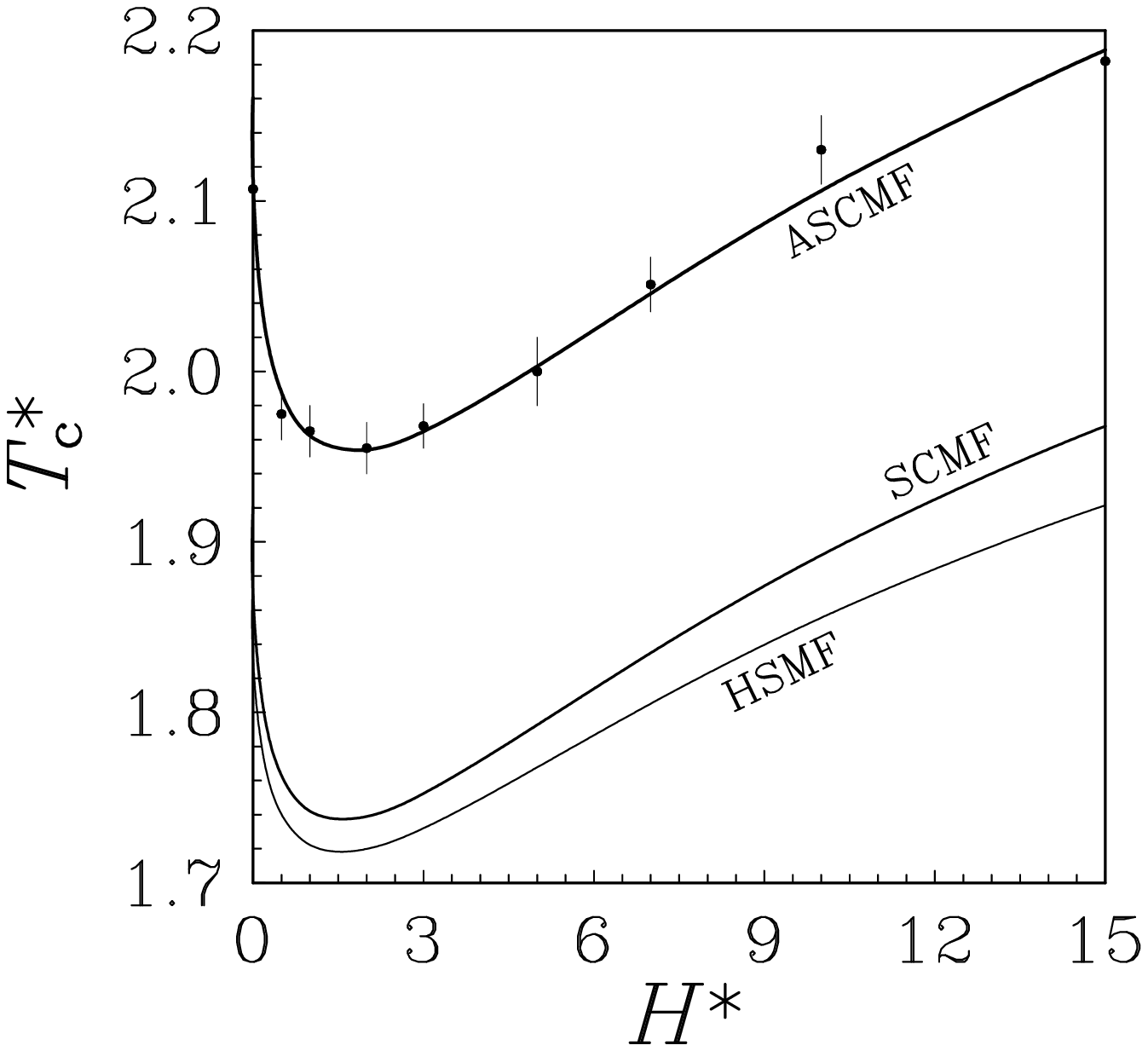}}

\vspace{6pt}

{\small FIG.~10. The same as in Fig.~9, but the behavior of $T_{\rm
c}(H)$ at small values of $H$ is shown in more detail.}
\end{figure}

\vspace{12pt}

For completeness of our consideration we present in detail in Fig.~11
the normalized dependencies $\rho_{\rm c}(H)/\rho_{\rm c\infty}$
of the critical density on the external field for the Heisenberg as
well as XY, and Ising fluids, obtained within the ASCMF theory in
comparison with the GEMC results. The normalization allows to make
a quite visible presentation of all the functions using only one
graph. As can be seen, the simulation points agree well with the
ASCMF predictions, although the deviations are somewhat larger than
in the case of the field dependency $T_{\rm c}(H)$ of the critical
temperature. The main part of these deviations should be associated
with MC statistical noise. For the absolute values of $\rho^\ast_{\rm
c\infty}=\lim_{H \to \infty} \rho^\ast_{\rm c}$, related to the ASCMF
theory and GEMC simulations, we have obtained 0.295 and 0.307,
respectively. The nonmonotonicity of $\rho_{\rm c}(H)$ for $n=3$
and 2 is closely connected with the corresponding nonmonotonous
behavior of $T_{\rm c}(H)$. With switching on the external field and
its slight increasing, the critical density begins to increase rapidly
for all $n=1$, 2, and 3. Further, the Heisenberg and XY dependencies
$\rho_{\rm c}(H)$ exhibit a maximum at $H^\ast \approx 2$, i.e.,
approximately at the same point, where the functions $T_{\rm c}(H)$
have a minimum (see Figs.~5 and 9). For $n=3$ and 2, after reaching
the maximum, the functions $\rho_{\rm c}(H)$ decrease tending to
the same limiting value $\rho_{\rm c\infty}$. On the other hand,
for $n=1$ the function $\rho_{\rm c}(H)$ continues to increase
monotonically to $\rho_{\rm c\infty}$ on the whole interval of
varying $H$ (at the same time, the function $T_{\rm c}(H)$
monotonically decreases with rising $H$, see Fig.~2).

\vspace{12pt}

\begin{figure}[htbp]
\epsfxsize=84mm
\centerline{\epsffile{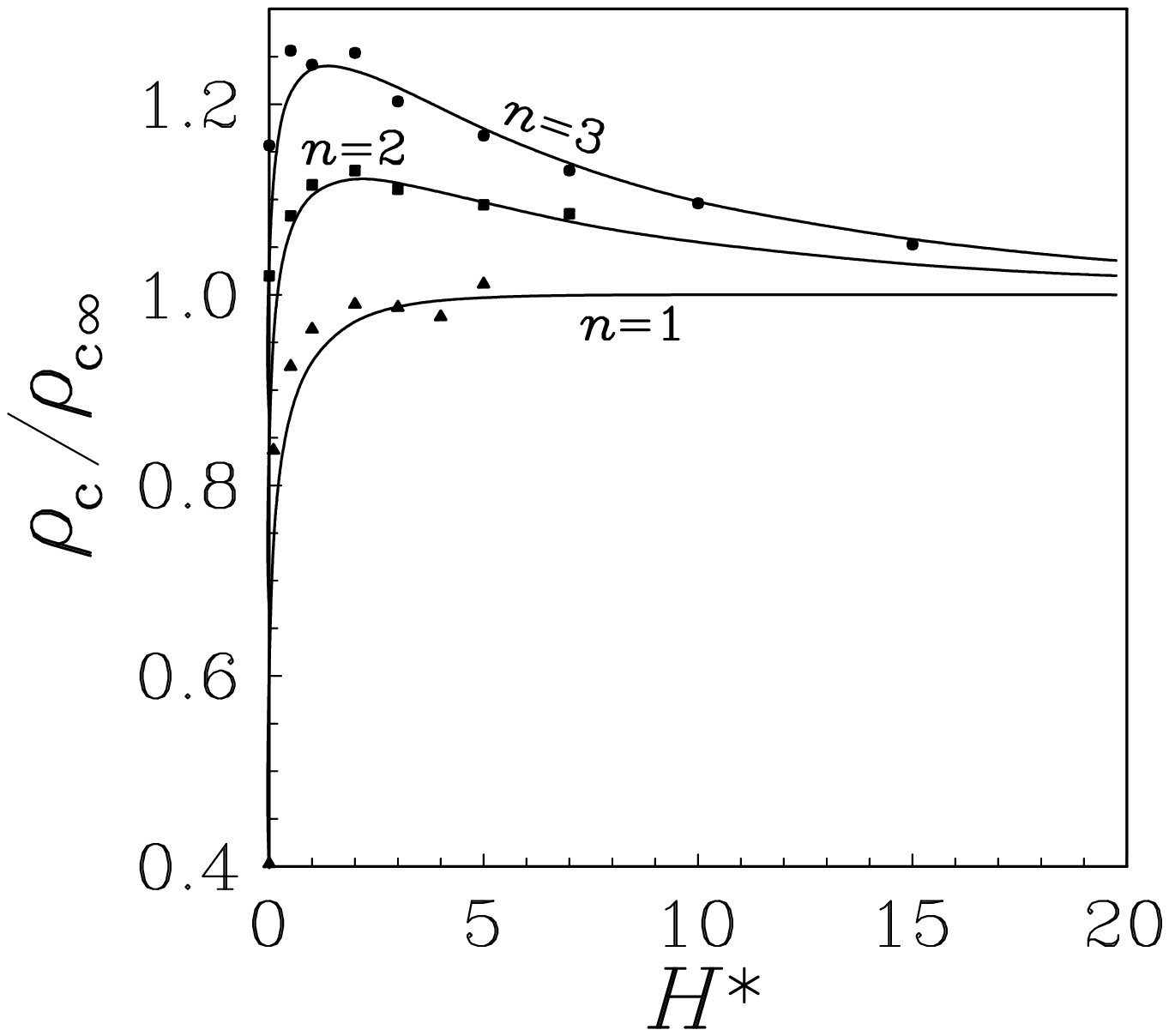}}

\vspace{6pt}

{\small FIG.~11. The normalized critical density $\rho_{\rm c}/
\rho_{\rm c\infty}$ as a function of the applied external magnetic
field $H$, evaluated within the ASCMF theory (curves) for the Ising
($n=1$), XY ($n=2$), and Heisenberg ($n=3$) spin fluid models. The
values obtained from GEMC simulations are presented by triangles,
squares, and circles, for $n=1$, 2, and 3, respectively.}
\end{figure}

\vspace{12pt}

Finally, the simulation and theory results on the para-ferro coexistence
are collected for $n=3$, 2, and 1 in Fig.~12. It repeats to some extent
the Curie lines already shown in Figs.~1, 3, 4, 6--8, but represents
them in a considerably wider temperature interval. Despite the fact that
only a set of three MC points (for each $n$) is available in a restricted
density region, it can be stated that the dependence of $T_\lambda$ on
$\rho$ is almost linear and excellently coincides with the MF straight
line $T_\lambda^\ast=\frac{8 \pi}{n} \rho^\ast$ (the MC uncertainties
are of order of the size of the symbols in Fig.~12).

\vspace{12pt}

\begin{figure}[htbp]
\epsfxsize=84mm
\centerline{\epsffile{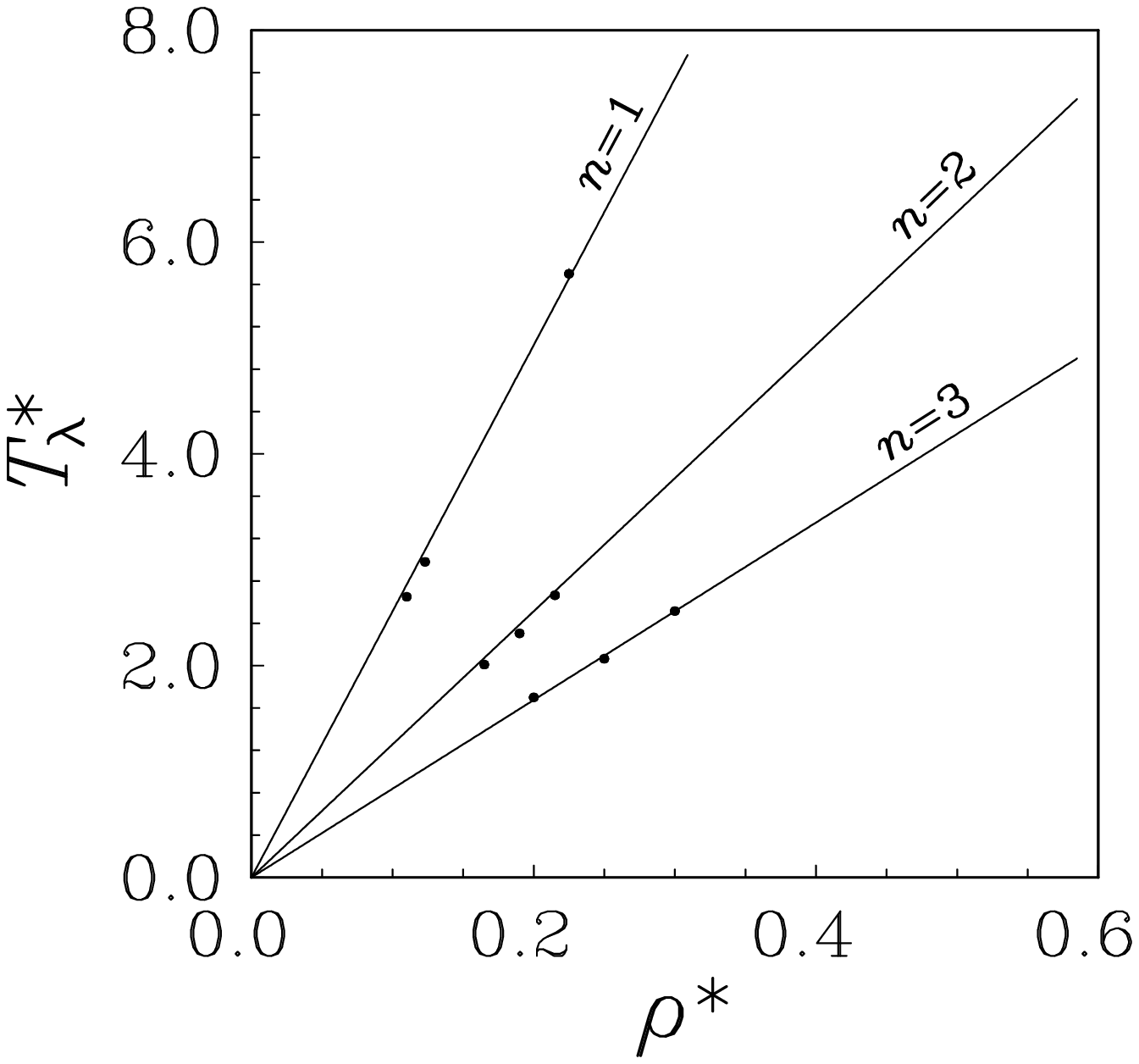}}

\vspace{4pt}

{\small FIG.~12. The para-ferro coexistence lines obtained within the
MF theory for the Ising ($n=1$), XY ($n=2$), and Heisenberg ($n=3$)
spin fluid models in comparison with canonical MC simulation data
(circles).}
\end{figure}

\vspace{6pt}

\section{Concluding remarks}

In the present study we have obtained a complete set of phase diagrams
for a class of Ising, XY, and Heisenberg ``ideal'' spin fluids. The phase
diagrams have been calculated applying a new version of the MF theory as
well as the Gibbs- and canonical-ensemble MC simulation techniques. The
new version takes into account the softness of nonmagnetic interparticle
repulsion and corrects the equation of state of the reference system.
This has allowed us to describe quantitatively the dependence of the
critical temperature $T_{\rm c}$ of the liquid-gas transition on
the strength $H$ of an external magnetic field for all the models
considered over the whole region of varying $H$. Let us summarize
and discuss the main results found:

(i) It has been established that a common feature inherent in the
function $T_{\rm c}(H)$ is its decrease with turning on and increasing
$H$. In the case of the Ising model ($n=1$), such a behavior remains
valid for any further increase of $H$, including the infinite field
regime ($H \to \infty$). For the XY and Heisenberg models ($n=2$ and
3), the decrease of $T_{\rm c}$ with rising $H$ gradually transforms
into the inverse dependence, i.e., into an increase of $T_{\rm c}$ at
intermediate and strong fields. In this respect the discrete Ising
model (where the spins can accept only two values, $+1$ or $-1$)
exhibits a very specific feature which is not observed for the XY
and Heisenberg models with continuous spin distributions.

(ii) From a physical point of view, the field effects just mentioned are
caused by the existence of two competing mechanisms. In order to make
the explanation more clear, one uses some expressions of the MF theory
(which describes already the main features of the field dependency of
$T_{\rm c}$ and $\rho_{\rm c}$). Then the spin fluid can be treated
as a simple nonmagnetic system with an effective attraction potential
$J_{\rm eff}(r)= -\langle {\bf s}_1 \cdot {\bf s}_2 \rangle_H J(r)$,
where $\langle {\bf s}_1 \cdot {\bf s}_2 \rangle_H=m^2$ within the MF
approach. This corresponds to the pressure $P=P_\varphi-a_{\rm eff}
\rho^2/2$ with $a_{\rm eff}=a m^2$ being the effective attraction
strength (see Eq.~(14)). Taking into account the dependence of $m$
on $\rho$ (and $H$), one obtains that $k_{\rm B} T_{\rm c} = \rho
a_{\rm eff} (1+a\rho \chi)/W'(\rho)$, where $\chi=\partial m/\partial
H$ is the magnetic susceptibility of the system. This result follows
from Eq.~(15), the condition $(\partial P/\partial \rho)_{T_{\rm c}}=
0$, and the MF relation $\partial m/\partial \rho= a m \chi$ (see
Eq.~(13)). Differentiating $T_{\rm c}$ with respect to $H$ yields
\begin{equation}
\frac{\partial T_{\rm c}}{\partial H} =
\frac{a m \rho}{k_{\rm B} W'} \left[ 2 \chi (1+a\rho\chi) +
a m \rho\frac{\partial \chi}{\partial H} \right] \, ,
\end{equation}
where the first and second terms on the right-hand side should be
associated with the contributions of two different mechanisms. The
first mechanism is due to the fact that the external field favors
to align the spins along ${\bf H}$ and thus increases $m$, i.e.,
$\chi > 0$ for arbitrary values of $H$. For genuine nonmagnetic
systems, where $a_{\rm eff}$ is independent of $\rho$, one obtains
the well-known result $k_{\rm B} T_{\rm c} = \rho a_{\rm eff}/W'$. This
obviously means that the critical temperature is higher for fluids with
stronger attractions between particles (e.g., the liquid-gas transition
disappears completely for systems with only repulsive interactions, when
$a_{\rm eff}=0$). In our case, the effective attractions $a_{\rm eff}=a
m^2$ will rise with increasing the magnetic field because of rising $m$,
and therefore the critical temperature will increase also (note that the
term $2\chi (1+a\rho\chi)$ remains always positive, see Eq.~(27)).

The second mechanism is more subtle and caused by the dependence of
$\chi$ on $H$. At strong enough fields ($H \gg k_{\rm B} T$ and $H \gg
a\rho$), when the magnetization $m$ is very close to its saturation value,
we obtain from Eq.~(25) that $\chi = \partial m/\partial H \sim \exp[-2H/
(k_{\rm B}T)]$ for $n=1$ as well as $\chi \sim 1/H^2$ for $n=2$ and 3.
Then $\partial \chi/\partial H|_{n=1} \sim -\exp[-2H/(k_{\rm B}T)]<0$ and
$\partial \chi/\partial H|_{n=2,3} \sim -1/H^3<0$. Thus for $n=2$ and 3,
the influence of the second term in the right-hand side of Eq.~(27) on the
value $\partial T_{\rm c}/\partial H$ can be neglected (this term is higher
order of smallness in $H$ with respect to the first contribution). In other
words, the first mechanism dominates and this indeed is observed for the XY
and Heisenberg models, where $T_{\rm c}$ increases with rising $H$ in the
large magnetic field regime. This is, however, not the case for the Ising
fluid, where the second term (being negative) appears to be the same order
in $H$ and even greater in magnitude than the first contribution, owing
to the exponential field dependency. Then the critical temperature will
decrease with increasing $H$. In particular, for $H \to \infty$, combining
the two terms gives $\lim_{H \to \infty} k_{\rm B} \partial T_{\rm c}/
\partial H = - 8 (W'_{\rm c\infty}-1) \exp[-2 H/(k_{\rm B} T_{\rm
c\infty})]<0$.

At weak magnetic fields $H \sim 0$, the second mechanism will prevail for
all the models. Following arguments presented in Ref.~\cite{Schinagl} at
$n=1$, it can be shown within the MF approximation that $\partial T_{\rm
c}/\partial H \sim - C(n) H^{-3/5} < 0$ for any $n=1$, 2, and 3, where
$C(n)>0$ is the $n$-dependent coefficient of the proportionality. In
order to better understand why the second mechanism suppresses the
liquid-gas separation, one rewrites the term $a m \rho \partial \chi/
\partial H$ in the equivalent form $\rho \partial \chi/\partial \rho -
a \rho \chi^2$. Then adding it to the first term $2\chi (1+a\rho\chi)$
results in $2\chi (1+\frac12 a\rho\chi)+\rho \partial \chi/\partial
\rho$. Thus, the negative contribution of the second term into the
derivative $\partial T_{\rm c}/\partial H$ directly follows from the
fact that the susceptibility of the gaseous phase is larger than that
of the coexistent dense liquid phase, i.e., $\partial \chi/\partial
\rho < 0$. This means that the magnetization of the gaseous phase grows
stronger, making the two phases more indistinguishable from one another.
This effect is very strong in the limit $H \to 0$, where the gas branch
of the binodal is very close to the Curie line (along which $\chi \to
\infty$). At the same time, the liquid branch quickly deviates from
this para-ferro transition line, leading to especially large negative
values of $\partial \chi/\partial \rho$ (with $\lim_{T \to T_{\rm c}}
\partial \chi/\partial \rho \to -\infty$ at $H \to 0$).

(iii) We see, therefore, that for the Ising fluid, the second mechanism
dominates (owing to the discrete nature of spin reorientations in this
case) at all values of $H$. This explains the monotonic decrease of
$T_{\rm c}$ with rising $H$ for $n=1$. With increasing the number $n$
of spin components to 2 and 3, the ability of the external field to
decrease the critical temperature becomes smaller, and the first
mechanism begins to prevail at larger values of $H$. This leads to
the nonmonotonic behavior of $T_{\rm c}(H)$ for the XY ($n=2$) and
Heisenberg ($n=3$) models. The behavior of $\rho_{\rm c}(H)$ can
also be explained by appealing to the interplay between the same
two competing processes.

(iv) It has been shown that the MF theory does not predict at $H=0$
the existence of a critical point and critical end point, found in
the simulations for $n=2$ and 3, but leads instead to a tricritical
behavior, observed in the simulations for $n=1$. In this context it,
first, should be pointed out that we dealt with a specific class
of spin fluids, where the attraction between particles is due to
ferromagnetic interactions. According to the general classification,
this corresponds to an infinite value of the ratio $R=\int g_\varphi(r)
J(r) {\rm d} {\bf r}\big/\int g_\varphi(r) \varphi_{\rm attr}(r) {\rm
d} {\bf r}$ of the integrated strengths of the magnetic interactions
and an attractive part of the nonmagnetic interactions, $R = \infty$
(because $\varphi_{\rm attr}=0$). For finite values of $R$, different
types of the phase diagram topology can be observed \cite{HemImb77,%
Schinagl} within the MF theory (at $R=0$ we come to the usual
nonmagnetic fluids). Secondly, for $R = \infty$ and $H=0$, we have
concluded that the question on the topology of phase diagrams is
very delicate, and requires more accurate theories and more
sophisticated simulation techniques. These and related topics
will be addressed in further investigations.

\section*{Acknowledgment}

Part of this work was supported by the Fonds zur F\"orderung der
wissenschaftlichen Forschung under Project No.~P15247. I.M. and
I.O. thank the Fundamental Researches State Fund of the Ministry
of Education and Science of Ukraine for support under Project
No. 02.07/00303.

\end{multicols}

\end{document}